\documentclass[twocolumn]{aastex7}
\usepackage{mathtools} 
\usepackage{csvsimple}
\usepackage{longtable}
\usepackage{footmisc}
\usepackage{mhchem}
\renewcommand{\thefootnote}{\arabic{footnote}}
\newcommand{\ang}{\text{\AA}}

\newcommand{\sini}{\sin i_\star}
\newcommand{\vsini}{v\sini}
\newcommand{\red}[1]{#1}

\makeatletter
\let\orig@maketitle\maketitle
\renewcommand{\maketitle}{%
  \orig@maketitle
  \setcounter{footnote}{0}%
  \renewcommand{\thefootnote}{\arabic{footnote}}%
}
\makeatother

\shorttitle{The KPF SURFS-UP Survey I}
\shortauthors{Householder et al.}
\graphicspath{{./}{}}
\usepackage{orcidlink}

\begin{document}

\title{The KPF SURFS-UP Survey I: Transmission Spectroscopy of WASP-76 b}

\author[0000-0002-5812-3236]{Aaron Householder}
\altaffiliation{NSF Graduate Research Fellow}
\affiliation{Department of Earth, Atmospheric and Planetary Sciences, Massachusetts Institute of Technology, Cambridge, MA 02139, USA}
\affil{Kavli Institute for Astrophysics and Space Research, Massachusetts Institute of Technology, Cambridge, MA 02139, USA}
\email{aaron593@mit.edu}

\author[0000-0002-8958-0683]{Fei Dai}
\affiliation{Institute for Astronomy, University of Hawai`i, 2680 Woodlawn Drive, Honolulu, HI, 96822, USA}
\affiliation{Division of Geological and Planetary Sciences,
1200 E California Blvd, Pasadena, CA, 91125, USA}
\affiliation{Department of Astronomy, California Institute of Technology, Pasadena, CA 91125, USA}
\email{fdai@hawaii.edu}

\author[0000-0002-3239-5989]{Aurora Kesseli}
\affiliation{IPAC, Mail Code 100-22, Caltech, 1200 E. California Boulevard, Pasadena, CA 91125, USA}
\email{aurorak@ipac.caltech.edu}

\author[0000-0001-8638-0320]{Andrew W. Howard}
\affiliation{Department of Astronomy, California Institute of Technology, Pasadena, CA 91125, USA}
\email{ahoward@caltech.edu}

\author[0000-0003-1312-9391]{Samuel Halverson}
\affiliation{Jet Propulsion Lab, Pasadena, CA 91125, USA}
\email{samuel.halverson@jpl.nasa.gov}

\author[0000-0003-3504-5316]{Benjamin J.\ Fulton}
\affiliation{NASA Exoplanet Science Institute/Caltech-IPAC, California Institute of Technology, Pasadena, CA
91125, USA}
\email{bjfulton@ipac.caltech.edu}

\author[0000-0003-0097-4414]{Yapeng Zhang}
\altaffiliation{51 Pegasi b fellow}
\affiliation{Department of Astronomy, California Institute of Technology, Pasadena, CA 91125, USA}
\email{yapzhang@caltech.edu}

\author[0000-0001-7047-8681]{Alex S. Polanski} 
\altaffiliation{Percival Lowell Fellow}
\affiliation{Department of Physics and Astronomy, University of Kansas, Lawrence, KS, USA}
\affiliation{Lowell Observatory, 1400 W Mars Hill Road, Flagstaff, AZ, 86001, USA}
\email{aspolanski@lowell.edu}

\author[0000-0001-9164-7966]{Julie Inglis}
\affil{Division of Geological and Planetary Sciences, California Institute of Technology, Pasadena, CA 91125, USA}
\email{jinglis@caltech.edu}

\author[0000-0001-9686-5890]{Nick Tusay}
\affiliation{Department of Astronomy, University of Washington, Seattle, WA, 98195, USA}
\email{ntusay@uw.edu}

\author[0000-0003-3355-1223]{Aaron Bello-Arufe}
\affiliation{Jet Propulsion Laboratory, California Institute of Technology, Pasadena, CA 91109, USA}
\email{aaron.bello.arufe@jpl.nasa.gov}

\author[0000-0002-6525-7013]{Ashley D.\ Baker}
\affiliation{Caltech Optical Observatories, California Institute of Technology, Pasadena, CA 91125, USA}
\email{abaker@caltech.edu}

\author[0000-0002-7226-836X]{Kevin B. Burdge}
\affiliation{Department of Physics, Massachusetts Institute of Technology, Cambridge, MA 02139, USA}
\affiliation{Kavli Institute for Astrophysics and Space Research, Massachusetts Institute of Technology, Cambridge, MA 02139, USA}
\email{kburdge@mit.edu}

\author[0009-0002-2419-8819]{Jerry Edelstein}
\affiliation{Space Sciences Laboratory, University of California, Berkeley, CA 94720, USA}
\email{jerrye@ssl.berkeley.edu}

\author[0000-0002-8965-3969]{Steven Giacalone}
\altaffiliation{NSF Astronomy and Astrophysics Postdoctoral Fellow}
\affiliation{Department of Astronomy, California Institute of Technology, Pasadena, CA 91125, USA}
\email{steveng@caltech.edu}

\author[0009-0004-4454-6053]{Steven R. Gibson}
\affiliation{Caltech Optical Observatories, Pasadena, CA, 91125, USA}
\email{sgibson@caltech.edu}

\author[0000-0003-0742-1660]{Gregory J. Gilbert}
\affiliation{Department of Astronomy, California Institute of Technology, Pasadena, CA 91125, USA}
\email{ggilbert@caltech.edu}

\author[0000-0002-9305-5101]{Luke B. Handley}
\altaffiliation{NSF Graduate Research Fellow}
\affiliation{Department of Astronomy, California Institute of Technology, Pasadena, CA 91125, USA}
\email{lhandley@caltech.edu}

\author[0000-0002-0531-1073]{Howard Isaacson}
\affiliation{501 Campbell Hall, University of California at Berkeley, Berkeley, CA 94720, USA}
\affiliation{Centre for Astrophysics, University of Southern Queensland, Toowoomba, QLD, Australia}
\email{hisaacson@berkeley.edu}

\author[0000-0003-2451-5482]{Russ R. Laher}
\affiliation{NASA Exoplanet Science Institute/Caltech-IPAC, 1200 E California Blvd, Pasadena, CA 91125, USA}
\email{laher@ipac.caltech.edu}

\author[0000-0003-0967-2893]{Erik A. Petigura}
\affiliation{Department of Physics \& Astronomy, University of California Los Angeles, Los Angeles, CA 90095, USA}
\email{petigura@astro.ucla.edu}

\author{Kodi Rider}
\affiliation{Space Sciences Laboratory, University of California, Berkeley, CA 94720, USA}
\email{kodi.rider@ssl.berkeley.edu}

\author[0000-0001-8127-5775]{Arpita Roy}
\affiliation{Astrophysics \& Space Institute, Schmidt Sciences, New York, NY 10011, USA}
\email{arpita308@gmail.com}

\author[0000-0003-3856-3143]{Ryan A. Rubenzahl}
\affiliation{Center for Computational Astrophysics, Flatiron Institute, 162 Fifth Avenue, New York, NY 10010, USA}
\email{rrubenzahl@flatironinstitute.org}

\author{Chris Smith}
\affiliation{Space Sciences Laboratory, University of California, Berkeley, CA 94720, USA}
\email{christopher.smith@berkeley.edu}

\author[0000-0001-7246-5438]{Andrew Vanderburg}
\affiliation{Center for Astrophysics | Harvard \& Smithsonian, 60 Garden Street, Cambridge, MA 02138, USA}
\email{avanderburg@cfa.harvard.edu}

\author[0000-0002-6092-8295]{Josh Walawender}
\affiliation{W.\ M.\ Keck Observatory, 65-1120 Mamalahoa Hwy, Kamuela, HI 96743, USA}
\email{jwalawender@keck.hawaii.edu}

\author[0000-0002-3725-3058]{Lauren M. Weiss}
\affiliation{Department of Physics and Astronomy, University of Notre Dame, Notre Dame, IN, 46556, USA}
\email{lweiss4@nd.edu}

\setcounter{footnote}{0}

\begin{abstract}
We introduce the KPF SURFS-UP (\textbf{S}pectroscopy of the \textbf{U}pper-atmospheres and \textbf{R}e\textbf{F}ractory \textbf{S}pecies in \textbf{U}ltra-hot \textbf{P}lanets) Survey, a high-resolution survey to investigate the atmospheric composition and dynamics of a sample of ultra-hot Jupiters with the Keck Planet Finder (KPF). Due to the unique design of KPF, we developed a publicly available pipeline for KPF that performs blaze removal, continuum normalization, order stitching, science spectra combination, telluric correction, and atmospheric detection via cross-correlation. As a first demonstration, we applied this pipeline to a transit of WASP-76 b and achieved some of the highest signal-to-noise detections of refractory species in WASP-76 b to date (e.g., Fe I is detected at a SNR of 14.5). We confirm previous observations of an asymmetry in Fe I absorption, but find no measurable ingress–egress asymmetry in Na I and Ca II. Together, these results suggest variations within different layers of the atmosphere of WASP-76 b: neutral metals such as Fe I trace deeper regions with stronger asymmetries, while Na I and Ca II probe regions higher in the atmosphere where the ingress-egress asymmetries are weaker. \red{Unlike some other ultra-hot Jupiters, our results are qualitatively consistent with GCM predictions of decreasing velocity asymmetry with altitude and do not require a high-altitude super-rotating jet that has been invoked for other planets (e.g., WASP-121 b). These results suggest that atmospheric circulation patterns in ultra-hot Jupiters may be more diverse than previously thought, highlighting the need for broader surveys to study how atmospheric dynamics depend on planetary and stellar properties.} 

\end{abstract}
\renewcommand{\thefootnote}{\arabic{footnote}}
\setcounter{footnote}{0}
\section{Introduction} 
Ultra-hot Jupiters (UHJs) are a rare class of gas giants with equilibrium temperatures greater than $\sim$2000-2200 K \citep{Parmentier2018, Stangret2022}. These extreme conditions make UHJs exceptional laboratories for studying atmospheric physics under unique temperature and pressure regimes. Indeed, many studies have used UHJs to observationally explore the three-dimensional nature of exoplanet atmospheres for the first time (e.g., \citealt{Ehrenreich2020, Tabernero2021, Gandhi2022, Kesseli2024, Beltz2024}).
In addition to atmospheric dynamics, UHJs also offer a unique window into planet formation and evolution. Their high temperatures mean that refractory species (e.g., Fe, Mg, Si) are often present in their atmospheres in gaseous form. This allows us to constrain refractory abundances, which may provide important clues about the formation location and migration history of hot Jupiters \citep{Lothringer2021,Chachan2023}. 

In this paper, we introduce the KPF SURFS-UP Survey to study UHJs using the recently commissioned Keck Planet Finder (KPF) on Keck I \citep{Gibson2016,Gibson2018,Gibson2020,Gibson2024, Sirk2018, Lilley2022,Rubenzahl2023}. KPF is a fiber-fed optical high-resolution (R $\sim$ 97,000; \citealt{Gibson2024}) spectrograph designed to achieve 30-50 cm s$^{-1}$ radial velocity precision, which is ideal for measuring small planet masses (e.g., \citealt{Dai2024}), low-amplitude Rossiter-McLaughlin (RM) signals (e.g., \citealt{Lubin2024}), and asteroseismology signals at the tens of cm s$^{-1}$ level (e.g., \citealt{Li2025}). The design of KPF also makes it one of the premier instruments for optical high-resolution atmospheric characterization of UHJs due to its high spectral resolution and exquisite line profile stability (e.g., a reloaded RM observation of KELT-18 b had a residual RMS of $\sim$185 $\pm$ 30 ppm; \citealt{Rubenzahl2024}). It also covers redder wavelengths (up to 8700 \ang) than many optical high-resolution spectrographs, which allows KPF to probe atomic and molecular species that absorb more strongly at redder wavelengths \red{(e.g., metal oxides and metal hydrides like TiO and FeH)}. Here, we present the first results of the KPF SURFS-UP Survey on the canonical UHJ WASP-76 b. In Section \ref{observations}, we describe our observations of WASP-76 b with KPF, and then discuss our pipeline to reduce the KPF data (Section \ref{reduction}). We then discuss our cross-correlation method (Section \ref{cc}), present the results of our atmospheric analysis (Section \ref{Results}), and place our findings in the broader context of prior high-resolution studies of WASP-76 b (Section \ref{discussion}). Finally, in Section \ref{conclusion}, we summarize our findings and outline future directions for the KPF SURFS-UP Survey.

\section{KPF Observations of WASP-76}
\label{observations}

On UT 2023-09-12, we observed a transit of WASP-76 b with KPF. Using the NASA Exoplanet Archive transit prediction service \citep{Christiansen2025} and the updated ephemeris from \citet{Kokori2023}, the transit midpoint was predicted to occur at UT 11:54:00 with a propagated uncertainty of 16 seconds. With an exposure time of 300 seconds, we obtained 60 observations of WASP-76, consisting of 34 in-transit and 26 out-of-transit exposures. Our first and last observations were taken at UT 09:14:05 and UT 15:09:27 (mid-exposure times), corresponding to airmasses of 1.71 and 1.29, respectively. We also took five ``slew calibrations'' approximately every hour to track instrumental drift throughout the observations (see Section \ref{wavecal} for details).

\begin{figure*}
    \centering
    \includegraphics[width=0.8\linewidth]{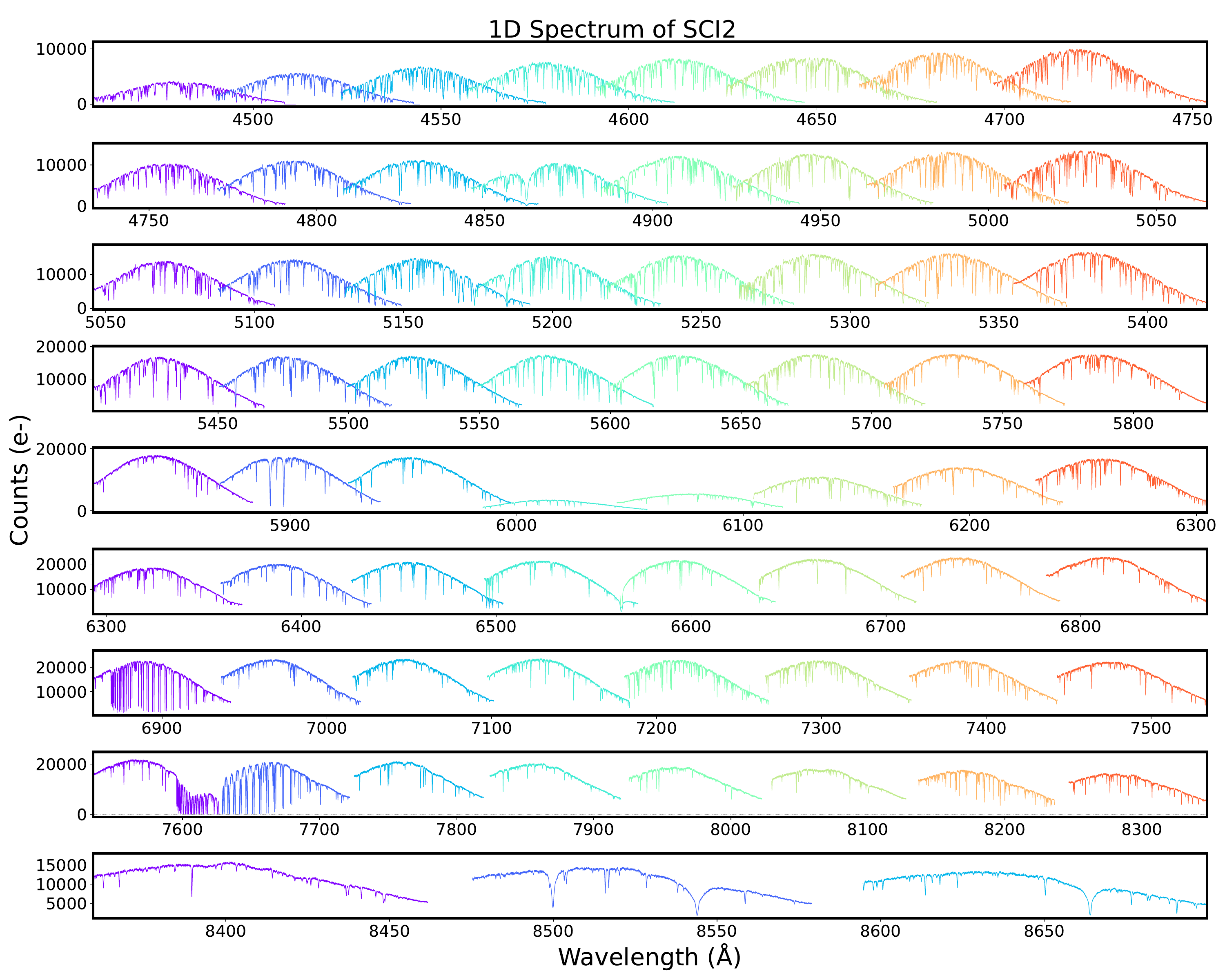}
    \caption{KPF spectra of WASP-76 extracted by version 2.7.1 of the KPF Data Reduction Pipeline. This plot shows only one of the three science spectra (SCI2), although all three spectra have similar looking spectra. Each color corresponds to a single diffraction order, and each row displays a distinct wavelength region for visual clarity. Before performing high-resolution atmospheric characterization, we performed several additional reduction steps, including blaze removal, continuum normalization, order stitching, combining the three science spectra, and telluric correction (see Section \ref{reduction}).}
    \label{spectrum}
\end{figure*}

Seeing conditions were favorable (hovering around 0.6$^{\prime\prime}$) for most of the night, but slightly deteriorated to $\sim$1.0$^{\prime\prime}$ towards the end of the observations. We examined all 60 exposures using diagnostics automatically produced by the KPF Quicklook Pipeline (based on signal-to-noise, guiding performance, etc). All observations passed these diagnostic checks except for the eighth exposure in the sequence, which was affected by a readout failure of the green CCD. We reduced the 59 good exposures with version 2.7.1 of the KPF Data Reduction Pipeline (DRP), an example of which is shown in Figure \ref{spectrum}. The 59 spectra have a median signal-to-noise ratio (SNR) of 255 near 7470 $\ang$ (where SNR is reported as the combined SNR of all three science spectra; see Section \ref{threespectra}). 

\section{Preparing the Spectra for High-Resolution Atmospheric Analysis}
\label{reduction}

\begin{figure}
    \centering
    \includegraphics[width=0.8\linewidth]{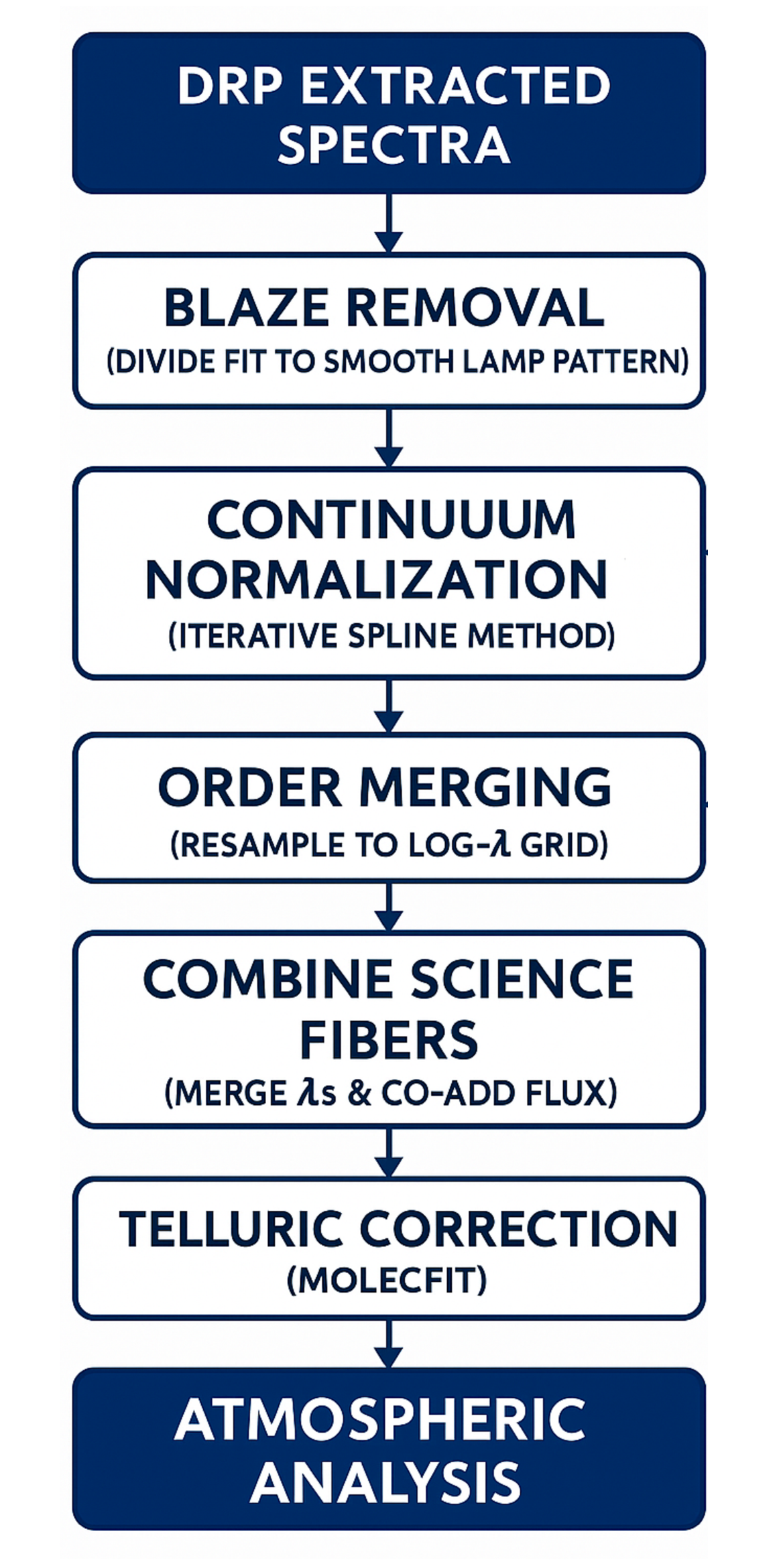}
    \caption{A general overview of the reduction steps for our pipeline. The three science spectra (SCI1, SCI2, and SCI3) are processed independently through blaze removal (Section \ref{sec:blaze-correction}), continuum normalization (Section \ref{sec:cont-norm}), and order merging (Section \ref{stitching}), and then combined (Section \ref{threespectra}) before telluric correction (Section \ref{tellurics}) and the subsequent atmospheric analysis (Section \ref{cc}). While the specific implementation of each step (e.g., continuum normalization) may evolve with future improvements, these steps will likely all remain important for preparing KPF data for atmospheric characterization.}
    \label{fig:reduction-steps}
\end{figure}

Before performing high-resolution transmission spectroscopy, we applied several additional steps to further reduce and calibrate the 1D extracted spectra. Here, we present our publicly available pipeline\footnote{https://github.com/aaronhouseholder/pancake}, which produces a single continuum-normalized, order combined, telluric-corrected spectrum, suitable for high-resolution atmospheric characterization and other science cases involving 1D spectra (e.g., stellar characterization). A summary schematic overview of the reduction process is shown in Figure~\ref{fig:reduction-steps}. We also provide our recommendations based on testing several strategies for each step of the reduction process. 

\subsection{The Design of KPF: Two CCDs and Three Science Spectra}
KPF uses both a green and a red CCD to optimize spectral coverage and maintain high precision across a broad range of wavelengths \citep{Gibson2024}. The green channel has wavelengths from $\sim4450 - 6000$ $\ang$, while the red channel covers $\sim6000 - 8700$ $\ang$. We treat the extracted 1D spectra from both chips as a single spectrum from 4450 to 8700~\ang, omitting only the first $\sim$10~\ang\ of the red chip ($\sim$5990–6000~\ang) due to low flux at the dichroic boundary between cameras. In addition to using two CCDs, KPF optically slices its primary science fiber into three narrower resolution elements arranged in a pseudo-slit to maximize spectral resolution while minimizing the physical footprint of the spectrometer. This means that, for each exposure, there are effectively three independent science spectra, each with its own unique wavelength and flux values. While merging the spectra from the green and red CCDs is relatively straightforward, co-adding the three science spectra requires much more attention to detail (see Section \ref{threespectra}). After testing multiple strategies, we ultimately chose to combine the three different science spectra immediately before telluric correction (Section \ref{tellurics}). However, other data reduction steps such as wavelength calibration, blaze removal, continuum normalization, and order stitching steps are performed separately for each of the three science spectra.

\subsection{Wavelength Calibration}
\label{wavecal}
Wavelength calibration is the process of assigning a wavelength to each pixel in the spectrum (see e.g., \citealt{Zhao2021}). While the KPF DRP automatically produces wavelength-calibrated spectra, there are important subtleties in deciding which wavelength solution to adopt. For standard operations, KPF uses three wavelength calibration sources: a Laser Frequency Comb (LFC), a Thorium-Argon (ThAr) lamp, and a Fabry–P{\'e}rot etalon. For our observations of WASP-76~b, we used the nearest LFC and ThAr taken before and after the science exposures, with polynomials fit to their peaks to derive wavelength solutions. This approach provided us with two wavelength solutions: one from the LFC and one from the ThAr. Generally, the LFC is a superior calibration source compared to the ThAr lamp because the LFC provides a far denser line source: for any given order, there is roughly an order of magnitude more LFC lines than ThAr lines. However, one drawback of the KPF LFC during early science operations was that it struggled to produce flux in the bluest orders due to various challenges in generating a coherent supercontinuum in this wavelength regime (primarily caused by complex quantum nonlinear optics effects). As a result, during the calibrations for our WASP-76 observations, the KPF LFC produced little to no flux in the 12 bluest orders on the green chip. In comparison, the ThAr lamp provided wavelength calibration coverage across all orders. We adopted the default approach of the KPF DRP and used an LFC wavelength solution for orders that had LFC flux and a ThAr wavelength solution for orders without any LFC flux. \red{In the spectral regions where both calibration sources were available, the LFC and ThAr wavelength solutions were in excellent agreement at the m~s$^{-1}$ level, far below the atmospheric velocity shifts measured in this work.}

\subsection{Blaze Correction}
\label{sec:blaze-correction}
For echelle spectrographs, the blaze function refers to how light is distributed across the echelle orders, with the greatest amount of light centered at the ``blaze peak'' \citep{Schroeder1970}. The current extraction algorithms in the KPF DRP preserve the blaze function in the 1D spectra. While removing the blaze function is not strictly required for high-resolution atmospheric characterization, we included blaze removal in our pipeline because it makes later steps (e.g., combining the three science spectra) much more straightforward. To blaze correct our spectra, we divided the 1D WASP-76 spectra by a fit to the 1D ``smooth lamp pattern,’’ which we extracted from a stack of 2D flat-field images. This stack consisted of 32 flat-field exposures taken before our observations that were combined to increase the SNR and average over white noise. We then applied a low-pass filter to generate the smooth lamp pattern by averaging this stack over 200 pixels in the dispersion direction and 1 pixel in the cross-dispersion direction, clipping 3$\sigma$ outliers in the process. Using this 2D stacked image, we extracted a 1D version of the smooth lamp with the KPF DRP. We also applied a \texttt{W{\={o}}tan} \texttt{rspline} fit \citep{Hippke2019} to smooth over any bad pixels that were present in the 1D smooth lamp pattern and divided our WASP-76 data by these fits. A potential improvement to this technique would be to interpolate the smooth lamp pattern over the course of the night between the evening and morning calibrations (similar to what was done with etalon for wavelength calibration in Section \ref{wavecal}). This is because the blaze function of KPF can vary slightly over time (e.g., due to changes in the polarization of starlight). However, given the exquisite stability of KPF, these changes are expected to be minimal on the timescale of our WASP-76 observations.

\subsection{Continuum Normalization}
\label{sec:cont-norm}
\begin{figure}
    \centering
    \includegraphics[width=1\linewidth]{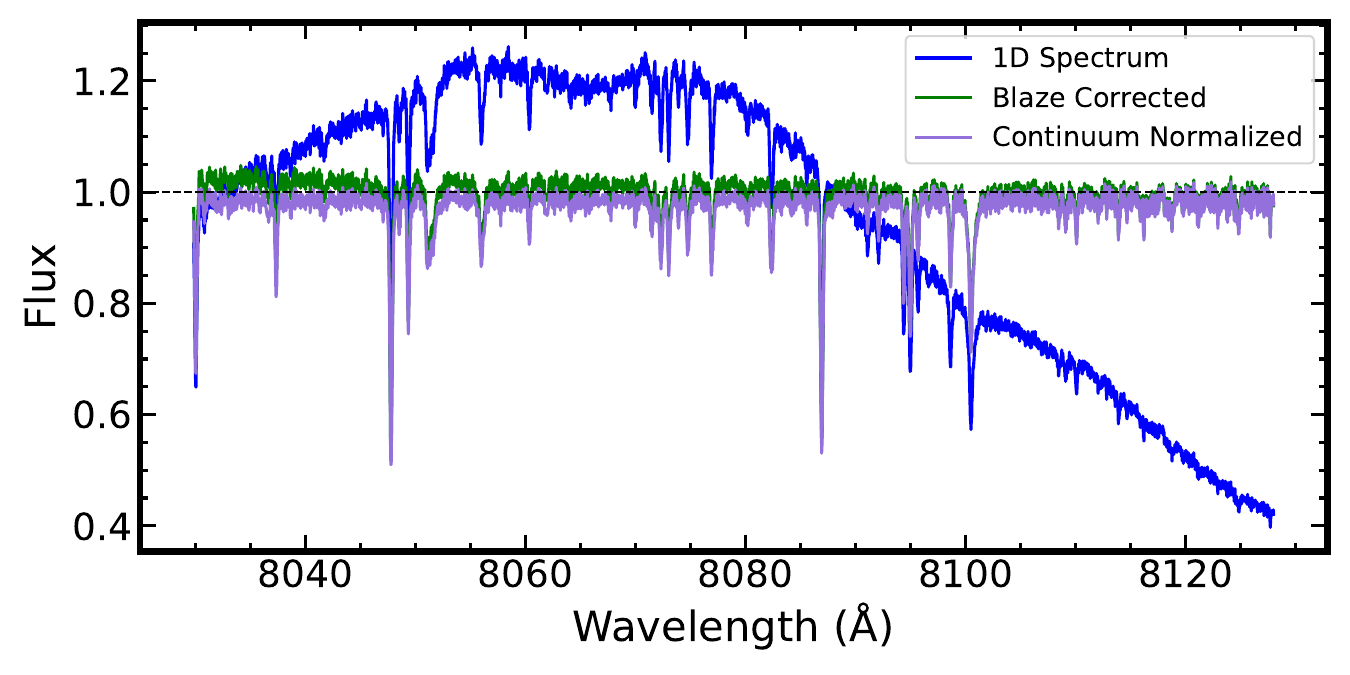}
    \caption{An example of the continuum normalization steps for a single order of a SCI2 spectrum of WASP-76 on the red CCD with significant fringing. First, the extracted 1D spectrum (blue) is divided by the smooth lamp pattern to produce the blaze-corrected spectrum (green). By comparing the blaze-corrected flux to a constant flux of 1.0 (top black dashed line), one can see a broad residual trend in the blaze corrected spectrum (i.e., the flux at bluer wavelengths is higher than the flux at redder wavelengths). To remove this trend, we implemented an iterative spline method on a master spectrum (constructed from the median of all exposures; see Section \ref{sec:cont-norm}) to produce the continuum normalized spectrum in purple.}
    \label{fig2}  
\end{figure}

The fit to the smooth lamp pattern that we divided out effectively represents the blaze function of KPF with the key difference that the smooth lamp pattern also includes the SED of the calibration lamp (rather than the SED of WASP-76). As a result, dividing by the fit to the smooth lamp pattern does not fully flatten the continuum for many orders (e.g., Figure \ref{fig2}). Across individual orders, we found that the trends in the residual continuum were roughly linear. However, some orders had more complex behavior, particularly toward the edges, where the intrinsic spectral dispersion can be highly non-linear. To correct these residual trends and normalize the continuum, we implemented an iterative spline fitting procedure. Instead of fitting splines to individual spectra (which could introduce systematics), we first created a ``master'' WASP-76 spectrum by calculating the median flux at each wavelength across all 59 observations of WASP-76.  We then fit an initial spline to each order on this master WASP-76 spectrum, determining the number of knots based on the wavelength range of each order. We used a window size of 10 \text{\AA} for almost all orders, except for a handful of orders where we increased the window size or chose not to continuum normalize due to extremely broad absorption features (e.g., H-alpha). After this step, we identified continuum points by calculating the ratio of the master spectrum to the spline and selecting regions where this ratio exceeded a threshold of 0.9875. \red{This threshold was chosen based on empirical testing, as values between 0.98 and 0.99 produced nearly identical results.}. We then refit the spline only to these continuum points to refine our estimate of the overall continuum. This process was repeated across all orders for five iterations or until convergence was achieved. We then divided the flux in our individual spectra by these continuum spline fits. As shown in Figure \ref{fig2}, this approach performs much better than simply dividing by the 1D smooth lamp pattern. For the last step of this continuum normalization process, we re-normalized the spectra in each order so that the 95th percentile of the flux was equal to 1,  following the approach of \citet{ValentiFischer2005}.  Importantly, high-resolution continuum normalization techniques continue to be an active area of research (e.g., \citealt{Rajpaul2020}), so future iterations of this pipeline may use more complex methods. However, based on some preliminary testing our spline-based approach worked much faster and produced similar  results compared to more complex approaches like alpha-shape fitting (e.g., \citealt{Xu2019}). 

At first glance, it may seem redundant to perform the blaze removal and continuum normalization steps separately. Indeed, for low-$\vsini$ stars like WASP-76~b the continuum normalization procedure alone should suffice to simultaneously remove the blaze and flatten the continuum. However, for stars with higher $\vsini$ (including most of the stars that host the UHJs in our sample), it becomes much more difficult to distinguish stellar absorption features from fringing \citep{Barden1998}, which is particularly prominent in the reddest KPF orders. Dividing by a fit to the smooth lamp pattern helps mitigate this degeneracy by removing low-frequency variations that are clearly not astrophysical in origin. After dividing the fit to the smooth lamp pattern, the remaining features in the spectra should primarily be stellar absorption, which makes it much easier to assess how well the continuum fit is performing. Although alternative approaches are possible (e.g., distinguishing fringing versus stellar absorption based on their frequency in cycles per angstrom), dividing by the smooth lamp pattern is a simpler solution. For this reason, we chose to perform the blaze correction and continuum normalization steps separately in our pipeline.

\subsection{Order Merging}
\label{stitching}
The combination of KPF gratings produces spectra that are dispersed across multiple diffraction orders. Although the diffraction orders overlap as they emerge from the grating, each order still contains a unique range of wavelengths that is not covered by adjacent orders (i.e., the free spectral range, or FSR, of an order). At shorter wavelengths, there is more order overlap because the FSR is inversely proportional to the order number that satisfies the diffraction grating equation \citep{Lowewen1997,Gibson2013}. For the same reason, at longer wavelengths, there is less order overlap as well as gaps in wavelength coverage. This behavior can be seen by eye in Figure \ref{spectrum}, and we opted to combine the orders to avoid handling overlapping wavelength regions and order-dependent weighting later in the pipeline, which simplifies the subsequent steps and cross-correlation analysis.

To combine the orders, we first removed wavelength regions with low SNR, using SNR values measured from the raw 1D spectra (assuming photon noise dominates over instrumental noise) rather than measuring the SNR from the continuum-normalized spectra. This preserves the true noise properties of the spectra, which are otherwise altered during blaze correction and normalization. We chose not to apply a single hard threshold when removing the data but rather inspected each order by eye to determine which wavelengths to remove. While deciding which regions to exclude can be somewhat subjective, almost all of the data we removed was towards the edges of the bluest orders where the SNR was lowest. Fortunately, because the bluest orders also have greater overlap (Figure \ref{spectrum}), removing them had no impact on wavelength coverage (i.e., the lower SNR edge of one blue order was covered by a higher SNR region of an adjacent order). 

After discarding low SNR wavelength regions, we stitched the orders together. We defined a uniform wavelength grid with a constant logarithmic step size, closely matching the average pixel size for each order.  A logarithmic grid is important because KPF reaches nearly constant resolving power across all wavelengths\footnote{This is not strictly true, as currently, the resolving power of KPF degrades toward the bluer wavelengths of the orders on the green CCD (Howard et al., in prep.), but we ignore this effect for simplicity.}, so one resolution element broadens as $\Delta\lambda\propto\lambda$. A fixed wavelength step in $\log \lambda$ follows the same scaling compared to a linear wavelength grid, which over-samples blue wavelengths (too many pixels per resolution element) and under-samples red wavelengths (too few pixels). For wavelength regions in the red with gaps in wavelength spacing between orders, we simply assigned a flux value of zero to ensure that each spectrum was continuous. After creating this wavelength grid, we then linearly interpolated the flux for all orders onto this grid to produce the final continuous spectrum for each of the science spectra. While we currently resample all spectra to a common wavelength grid, it may be possible to avoid any interpolation-induced correlations by fitting all exposures simultaneously in pixel space using a forward-model approach (e.g., \citealt{Hogg2024}). 

\subsection{Combination of the Three Separate Science Spectra}
\label{threespectra}

KPF has a large optical fiber (1.14'' in diameter on sky) and is located on a telescope with a large primary mirror (10 meters), making it well-suited for high photon throughput compared to many instruments on smaller telescopes. However, a key challenge of this design is to accept the large beam from the telescope and achieve a high spectral resolving power without scaling the dimensions of the spectrograph to an impractical size. To overcome this issue, KPF was designed with a reformatter, which rearranges light from the main science fiber into three narrower resolution elements (referred to as SCI1, SCI2, and SCI3). This means that, for each observation, KPF produces three separate science spectra, each with its own wavelength solution. Importantly, this design allows KPF to achieve high resolving power (R $\sim$ 97,000; \citealt{Gibson2024}) with a much smaller spectrometer, but it introduces several complexities that must be addressed when using KPF data. In particular, one consequence of using a reformatter is that it produces differences in the dispersion direction between SCI1, SCI2, and SCI3, so there are wavelength offsets between the spectra. A secondary wavelength solution effect comes from the fact that SCI1, SCI2, and SCI3 are aligned in a pseudo-slit that is not exactly vertical relative to the CCD pixels (see Figures 11 and 12 of \citealt{Gibson2024}). The angle between the CCD pixels and the pseudo-slit changes slightly from the left to right of the echelle orders, which creates an approximately constant offset in the wavelength solutions between the three spectra. Most of these wavelength differences are modeled out by wavelength calibration (Section \ref{wavecal}), but each spectrum still has its own unique wavelengths that must be merged or combined in some way. In addition to these wavelength differences, the spectral line shapes also vary between the three science spectra. This is because the primary science fiber is sliced in such a way that SCI2 has a nearly rectangular shape, while SCI1 and SCI3 have a trapezoidal-like shape (Figure \ref{fig:lfc_nad}). These different shapes mean that SCI1, SCI2 and SCI3 have different point spread functions (PSFs)/line spread functions (LSFs), which cause the line shapes between them to differ. 

Given these differences in wavelength and line shape, we explored multiple approaches for combining the three spectra in a way that preserves the fidelity of the atmospheric signal. Below, we discuss each strategy we tried, as well as the advantages and disadvantages of each method:

\begin{figure}
    \centering
    \includegraphics[width=0.5\textwidth]{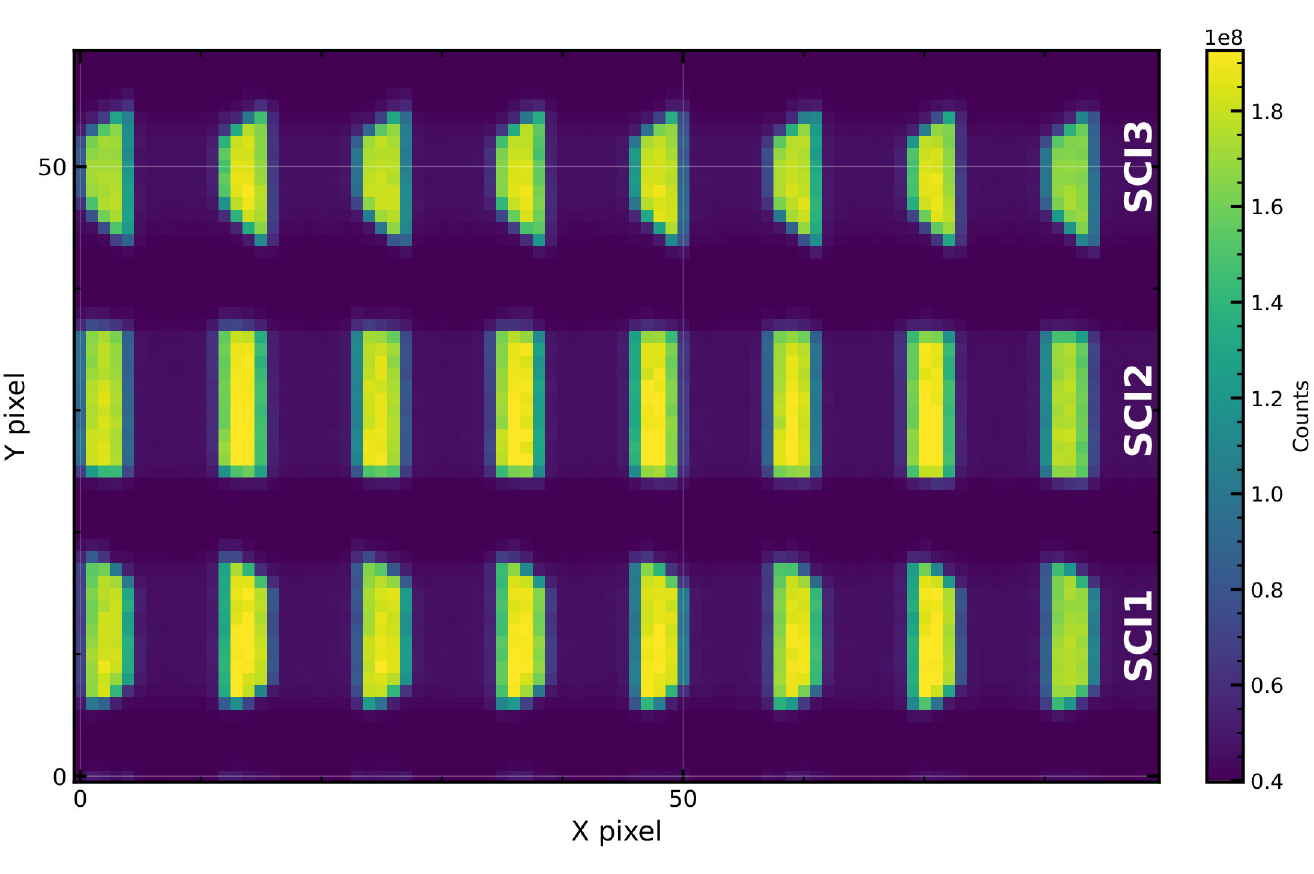}
    \caption{A zoom-in to a 2D laser frequency comb (LFC) frame to highlight the different PSFs between each of the three resolution elements. }
    \label{fig:lfc_nad}
\end{figure}

\begin{enumerate}
\item[(i)] \textbf{Wavelength Interpolation and Flux Co-addition:} Our first approach was to simply interpolate each of the three science spectra onto a common logarithmic grid and co-add the flux of the three spectra together. We tried this approach because the residual wavelength differences between SCI1, SCI2, and SCI3 were small (on the order of $10^{-1}-10^{-2} \text{\AA}$), so we did not expect interpolation on to a common wavelength grid to lead to significant errors. Furthermore, since wavelength interpolation was already required for order merging (Section \ref{stitching}), we did not expect this step to significantly affect the atmospheric analysis.

\item[(ii)] \textbf{Treating Each Spectrum Independently:} To test whether combining the spectra as in strategy (i) compromised our analysis, our next approach kept each spectra separate through telluric correction and cross-correlation. Only at the final stage did we co-add the CCFs (which are, by construction, produced on a common velocity grid) to produce a combined atmospheric detection signal.

\item[(iii)]\textbf{Forward‑modeling the LSFs:} 
This last method followed the procedure as in (ii), but before performing cross-correlation we wanted to account for the differences in line shape between the spectra. To do this, we modeled the LSF of each science spectra using the LFC exposures taken before our observations. The LFC exposures were a natural choice to characterize the instrumental profile of KPF due to their intrinsically narrow lines. For all three science spectra, we fit each LFC line with a Gaussian convolved with a top‑hat using \texttt{scipy.optimize.curve\_fit} \citep{Virtanen2020}. Handling the order overlap in the same way as in Section \ref{stitching}, this provided us with a wavelength‑dependent LSF for each science spectra. Each atmospheric template (see Section \ref{template}) was then convolved with this LSF model prior to performing cross-correlation transmission spectroscopy. 

\end{enumerate}

For each of the three spectral combination strategies described above, we repeated the same telluric correction steps and atmospheric analysis (Sections \ref{tellurics}, \ref{cc}, and \ref{Results}). Across all three methods, the detected signals were highly consistent, with no obvious trends between the methods and differences in detection significance of $<1.3$ in SNR for all detected species (see Section \ref{Results} for how significances were computed). Although the results were similar, we adopted strategy (i) as the default approach in our pipeline because it yielded more accurate telluric corrections (see Section \ref{tellurics}) and was more computationally efficient than strategy (ii). Although strategy (iii) allowed for explicit modeling of the spectra-dependent LSFs using the LFC, we found that the added complexity did not meaningfully improve the atmospheric detections. Because building these LSF models with the LFC takes considerable effort and did not produce noticeably better results for WASP-76 b, we therefore chose strategy (i) as the default approach in our pipeline. 

We stress that this simple interpolation and co-addition approach works well for transmission spectroscopy but may not be applicable for other science cases. For example, we caution against using a similar wavelength combination strategy for extreme precision radial velocity (EPRV) science, where sub-meter per second stability is required. In such cases, it is probably preferable to solve for RVs for SCI1, SCI2, and SCI3 independently. This technique would also likely not be ideal for measuring precise stellar properties such as macroturbulence or projected rotation velocities ($\vsini$), which are extremely sensitive to subtle changes in the LSF \citep{ValentiFischer2005,Doyle2014}. For such cases, we recommend using a similar approach to the LFC-based LSF correction outlined above or non-parametric approaches (e.g., \citealt{Schmidt2024}). \red{It is also likely possible to super-sample the spectra using the three independent resolution elements, since each resolution element provides a slightly different sampling of the PSF. However, we did not attempt this given the substantially different PSFs/LSFs of SCI1, SCI2, and SCI3.}

\subsection{Telluric Correction with \texttt{Molecfit}}
\label{tellurics}
After combining the science spectra together, the next step in our reduction was to correct the WASP-76 spectra for telluric absorption in Earth’s atmosphere. To do this, we used version 1.5.9 of \texttt{Molecfit} \citep{Smette2015}, which uses a line-by-line radiative transfer model to derive the transmission spectrum of the atmosphere across specific wavelength regions. We set up \texttt{Molecfit} to fit both \ce{H2O} and \ce{O2} lines, which are the dominant telluric absorbers in the KPF bandpass. Although \texttt{Molecfit} is effective for mitigating contamination from weak \ce{H2O} absorption lines, it can struggle with deeper absorption lines, such as the absorption in the cores of strong \ce{O2} lines, so we excluded those deeper absorption regions from the fit. We used vacuum wavelengths throughout, since that is the default output of the KPF DRP, and for all other parameters, we adopted the same \texttt{Molecfit} configuration as \citet{Allart2017}. As is discussed in Section \ref{threespectra}, we also experimented with running \texttt{Molecfit} on each individual science spectra before combining them, but we found that correcting the combined spectrum for tellurics produced better fits in terms of the RMS of the residuals. Given the importance of precise telluric correction in high-resolution transmission spectroscopy \citep[e.g.,][]{Langeveld2021}, this was part of the reason we chose to use the combined-spectrum approach for our analysis. 

\begin{figure*}
    \centering
    \includegraphics[width=0.9\linewidth]{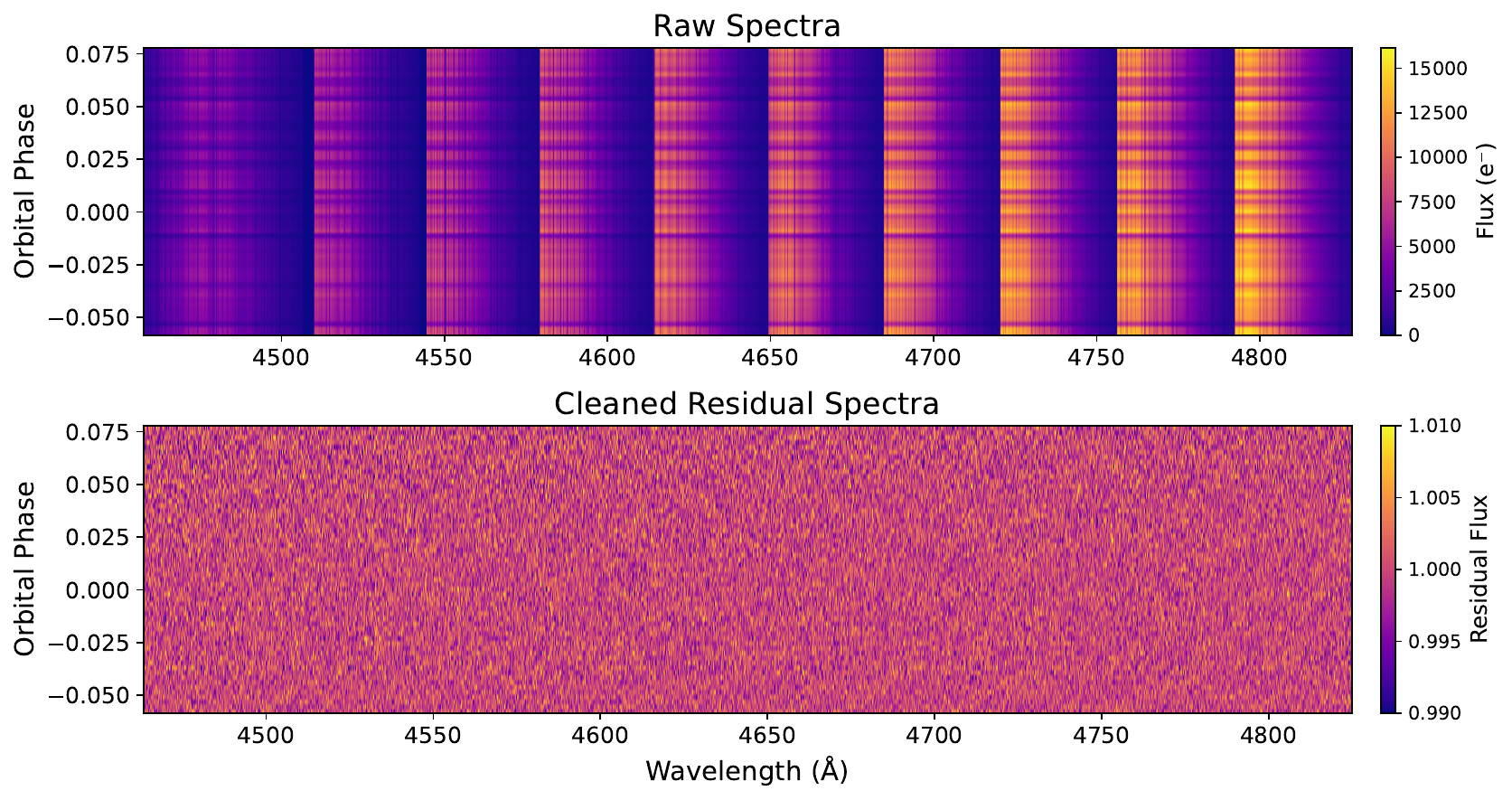}
    \caption{The results of our reduction pipeline for the first ten orders on the green chip. The top panel displays the raw extracted spectra from SCI2 as a function of the orbital phase of WASP-76 b, with each row representing a different exposure in the stellar rest frame. The bottom panel shows the same wavelength coverage after applying our reduction pipeline and dividing the flux by the out-of-transit average. These residuals contain the atmospheric absorption signal of WASP-76 b, which can be extracted using cross-correlation.}
    \label{fig:cleaning-results}
\end{figure*}

\subsection{Additional Cleaning and PCA}
\label{PCA}
During visual inspection of each spectrum, we noticed occasional pixel-scale outliers (e.g., isolated bad pixels with spuriously large flux). Future improvements to the DRP should reduce these issues, but for our analysis we applied an additional round of cleaning to remove these artifacts. We first masked wavelength bins where the normalized time-averaged flux fell below 0.05, and then re-normalized each exposure by its mean flux. \red{We then excluded wavelength bins with large temporal variability and removed isolated outlier points by applying 3$\sigma$ clipping to the flux values at each wavelength.} We experimented with running this cleaning step on the three individual spectra as well as on the combined spectrum, but the downstream atmospheric results were nearly indistinguishable in both cases. 

\red{Even after these additional cleaning steps, we observed systematic residuals that appeared both in and out of transit. Similar features have been independently seen in multiple KPF datasets and reduction pipelines (e.g., Tusay \& Kesseli et al. in prep.), suggesting they are instrumental in origin rather than astrophysical. We attribute these residuals to instrumental line profile variations over the course of the night. The LFC and ThAr calibrations also showed a drift of $\sim$10~m~s$^{-1}$ over the night. We tested linearly interpolating the wavelength solution between bracketing calibrations using intermediate etalon frames, which reduced the residuals but did not fully remove them. This suggests that the variations were not a pure Doppler shift but involved changes to the instrumental line profile itself. To account for these variations, we applied principal component analysis (PCA) to remove correlated structure from our spectral time series.} Given the 2D spectral matrix X ($m$ exposures $\times$ $n$ wavelength bins), we performed singular value decomposition X = USV$^T$, where the columns $u_i$ of U are the left singular vectors, S is a diagonal matrix containing the singular values $\sigma_i$ ordered from largest to smallest, and the columns $v_i$ of V are the right singular vectors. We then reconstructed the cleaned spectra by zeroing the first $k$ singular values:

\begin{equation}
X_{\text{cleaned}} = \sum_{i=k+1}^{r} \sigma_i u_i v_i^T + 1
\end{equation}
After some experimentation, we found that removing 3 principal components ($k = 3$) best suppressed these systematic features while preserving the planetary signal. The absorption features from WASP-76 b were not strongly affected because they rapidly change with the orbital velocity of the planet, whereas the dominant systematic modes remained much more stationary in the stellar rest frame, making the two largely orthogonal in the PCA basis. \red{We emphasize that the choice of correction method had no discernible impact on the final atmospheric signals reported in this work. We also note that this additional cleaning step may not be necessary in the future if upgrades to the KPF DRP reduce the instrumental profile variations responsible for these artifacts.}

\section{Cross-Correlation Search for Atmospheric Absorption}
\label{cc}
With our cleaned spectra, we began searching for absorption features in the atmosphere of WASP-76 b. For transmission spectroscopy, this process typically involves normalizing the entire spectral time series by a ``master'' spectrum taken outside of transit. The resulting in-transit residuals contain the planetary atmospheric signal (Figure \ref{fig:cleaning-results}), which is then cross-correlated with a high-resolution atmospheric template to identify specific atomic and molecular species. Here, we discuss the details of our cross-correlation method for transmission spectroscopy of WASP-76 b. 

\subsection{Atmospheric Template and Cross-Correlation}

\label{sec4.1}
As is referenced above, for cross-correlation transmission spectroscopy, it is necessary to create a high-fidelity atmospheric template that reflects the expected properties of the planetary atmosphere. To do this, we used \texttt{petitRADTRANS} \citep{Molliere2019}, which generates high-resolution spectra based on the atmospheric properties of the planet (pressure-temperature profile, chemical abundances, cloud properties, etc). For the planet properties of WASP-76 b, we used the parameters from \citet{Ehrenreich2020}. Following \citet{Kesseli2022}, we assumed an isothermal P-T profile with an equilibrium temperature of 3000 K. We added a gray cloud at 0.01 bars and included collision‐induced absorption (CIA) from \ce{H2}-\ce{H2} and \ce{H2}-\ce{He} to ensure a realistic continuum opacity. With this setup, we created templates at a resolution of $10^6$ for the following species: Ca II, CaH, Cr I, Fe I, Fe II, FeH, K I, Mg I, Na I, Si I, TiO, and V I \citep{Kurucz1979,Burrows2003,Mckemmish2024}. \red{We selected these species to compare KPF with previous high-SNR detections of WASP-76 b from ESPRESSO/HARPS-N, while also taking advantage of the redder KPF bandpass to search for species that have stronger absorption at redder wavelengths, such as FeH and TiO.} Abundances for individual species were determined using \texttt{easyCHEM} \citep{Molliere2017,Lei2024UpdateLater}, which computes abundances in exoplanetary atmospheres assuming chemical equilibrium.  To tailor these templates for cross-correlation, \red{we used Gaussian convolution to degrade the templates to R $=$ 97,000} and fit and subtracted a fourth-order polynomial to remove any residual continuum. We then normalized the resulting templates so that the sum of their flux was equal to unity. 

To perform the cross-correlation, we first placed each spectrum in the stellar rest frame using the best fit parameters from \citet{Ehrenreich2020} by correcting for the barycentric velocity ($v_{\rm bary}$), the systemic velocity ($v_{\rm sys}$), and the stellar reflex motion ($v_{\rm reflex}$): 
\begin{equation}
v_\star(t) = v_{\rm sys} - v_{\rm bary}(t) - v_{\rm reflex}(t).
\end{equation}
We then created a velocity grid from $-300$ to $+300$ km s$^{-1}$ in steps of 0.5 km s$^{-1}$. For each velocity shift, we Doppler-shifted the atmospheric template and linearly interpolated it onto the observed wavelength grid before computing the cross-correlation via the following equation:

\begin{equation}
c(\nu, t) = \sum_{i=0}^N x_i(t) T_i(\nu),
\label{eq1}
\end{equation}
where $c(\nu, t)$ is the two-dimensional cross-correlation function as a function of trial radial velocity $\nu$ and observation time $t$, $x_i(t)$ is the observed flux, and $T_i(\nu)$ is the template spectrum. After obtaining the 2D CCF grid, we normalized each CCF by the mean out-of-transit CCF to search for atmospheric absorption in-transit. For visualization purposes, we shifted each CCF into the planet rest frame using the expected orbital velocity $v_p(t)=K_p \sin(2\pi\phi(t))$. It is important to note that when using a synthetic model atmosphere, the reported residual amplitude (i.e., the strength of the detected signal in ppm) strongly depends on the assumed model parameters used to generate the template. \red{For instance, we found that varying the assumed atmospheric temperature used to generate the template from 2200 K (close to the equilibrium temperature of WASP-76 b) to 3000 K changed the recovered signal amplitudes by several hundred parts-per-million (ppm).} We therefore caution against interpreting the amplitudes as direct tracers of relative abundance or atmospheric altitude, since they are sensitive to the assumed model parameters used to generate each template.

\label{template} 

\section{Results: Atmospheric Signals}
\label{Results}

\subsection{Blue-shifted and Asymmetric Absorption}
\begin{figure*}[!ht]
    \includegraphics[width=0.49\linewidth]{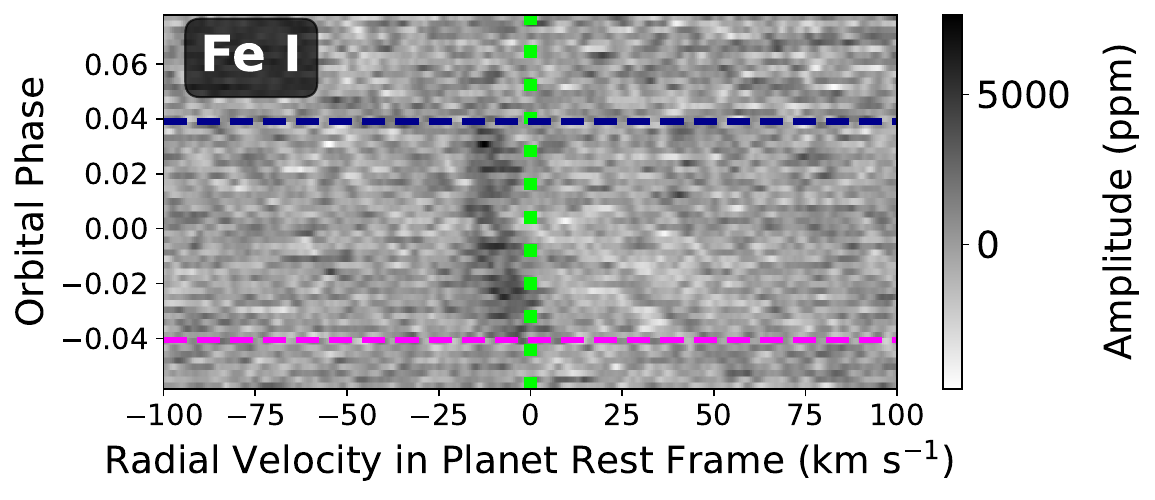}
    \includegraphics[width=0.49\linewidth]{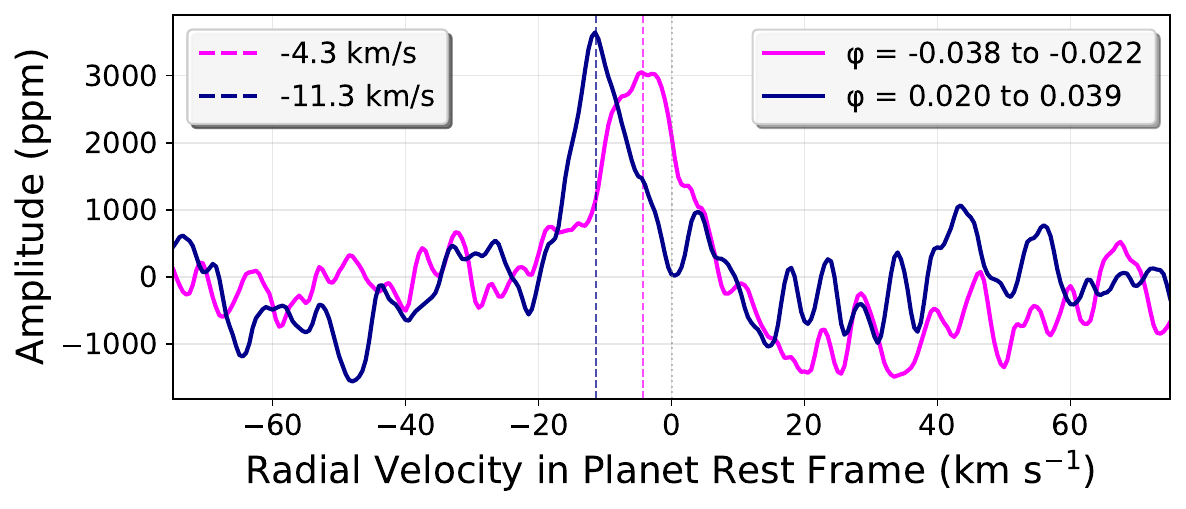}
    \includegraphics[width=0.49\linewidth]{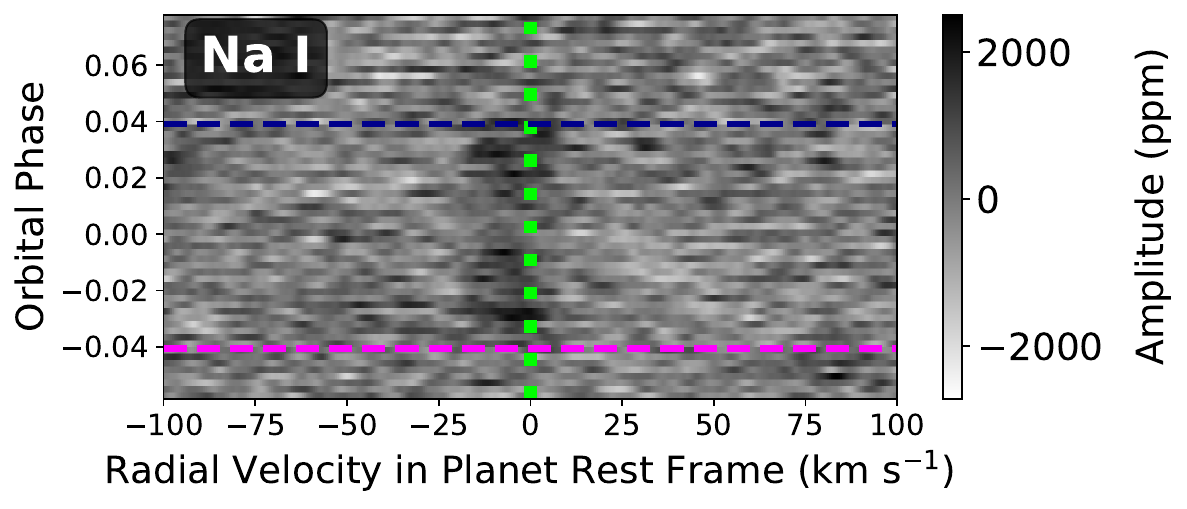} 
    \includegraphics[width=0.49\linewidth]{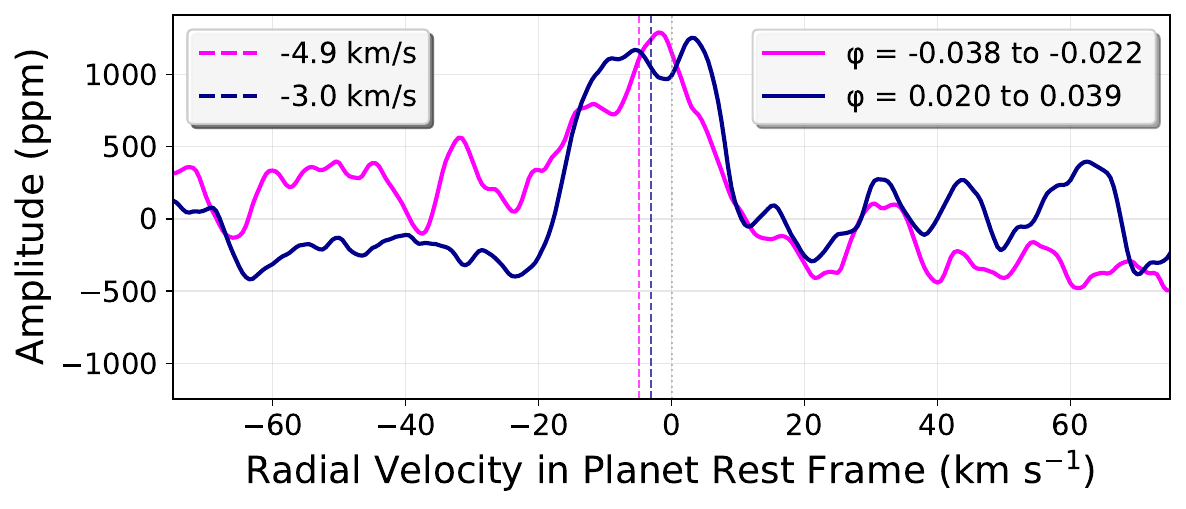}   
    \includegraphics[width=0.49\linewidth]{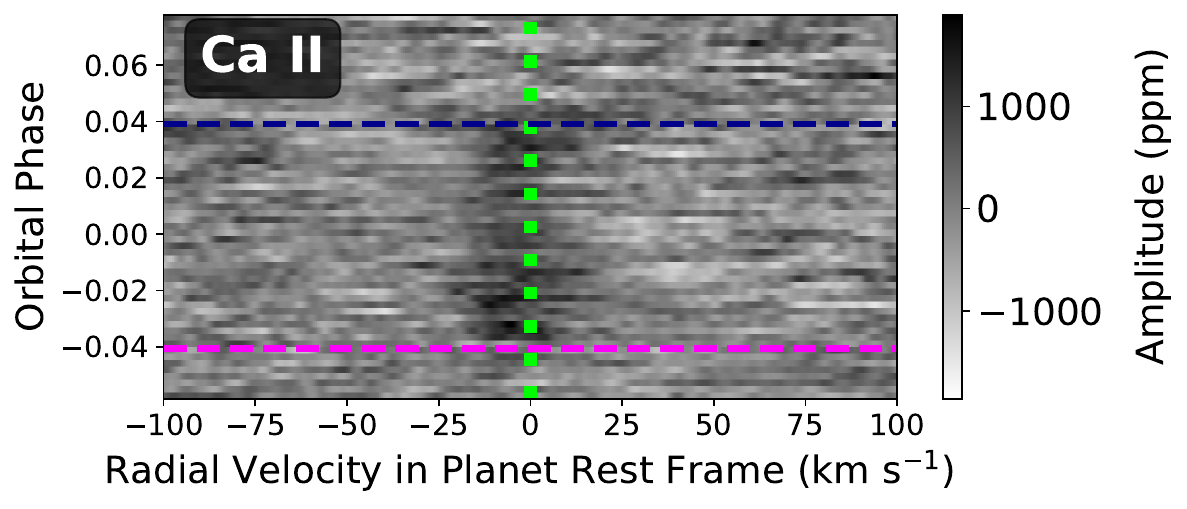} 
    \includegraphics[width=0.49\linewidth]{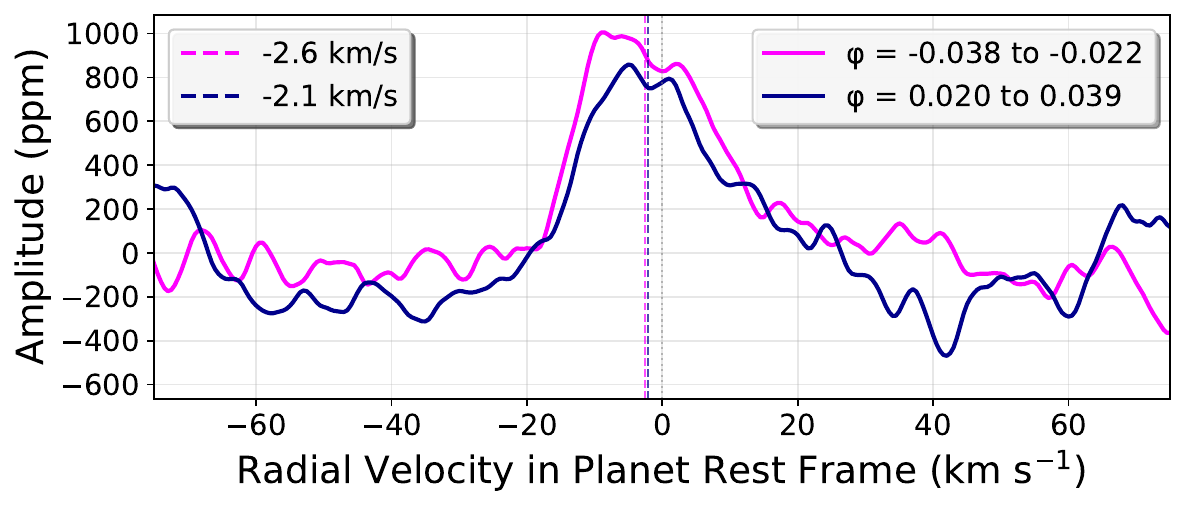}   
    \includegraphics[width=0.49\linewidth]{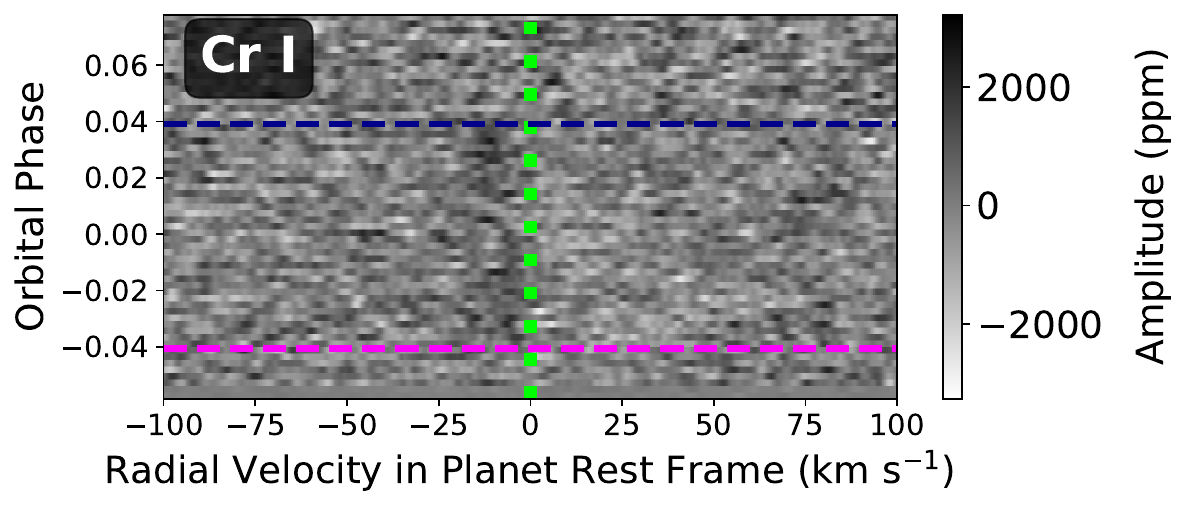} 
    \includegraphics[width=0.49\linewidth]{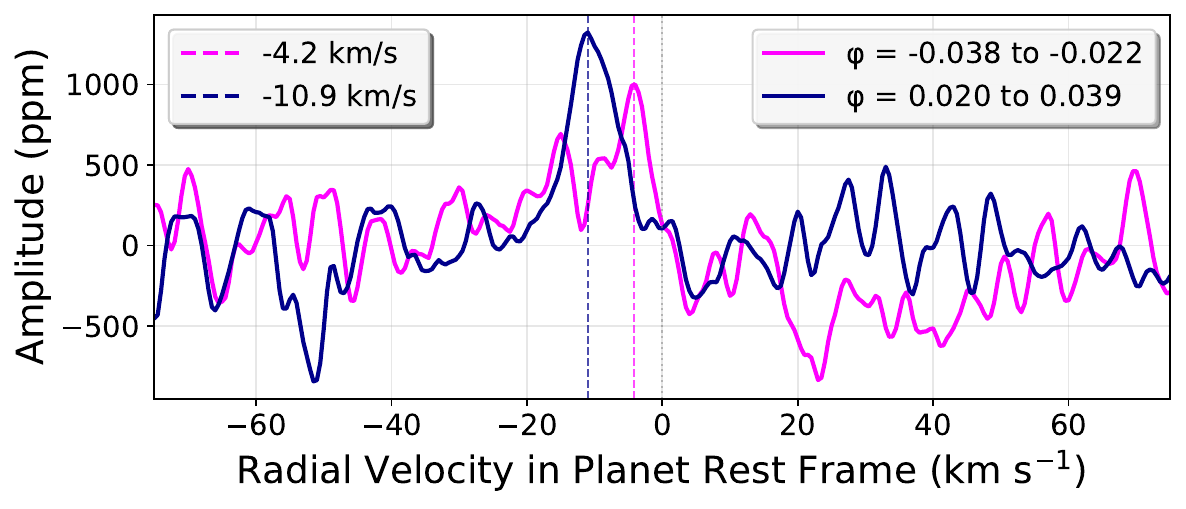} 
    
    \centering
    \caption{Left: Residual cross-correlation map for Fe I, Cr I, Na I, and Ca II from a single KPF transit of WASP-76 b, shown in the planet rest frame (0 km s$^{-1}$ is the green dashed line). The signal (in ppm) is plotted as a function of orbital phase (vertical axis) and radial velocity (horizontal axis), with each cross-correlation function (CCF) normalized by the out-of-transit average. The horizontal dashed lines mark the start of ingress (pink) and the end of egress (dark blue). In the absence of atmospheric dynamics, the planetary signal would be centered exactly at 0 km s$^{-1}$. Right: Fe I shows a clear ingress–egress asymmetry, while Na I and Ca II remain nearly vertical in velocity space with no measurable asymmetry. This difference points to altitude-dependent circulation differences: Fe I samples regions with stronger day-to-night winds, whereas Na I and Ca II probe higher altitudes where winds are weaker.}
    \label{fig:ccmap}
\end{figure*}

\begin{figure*}
    \centering
    \includegraphics[width=0.32\linewidth]{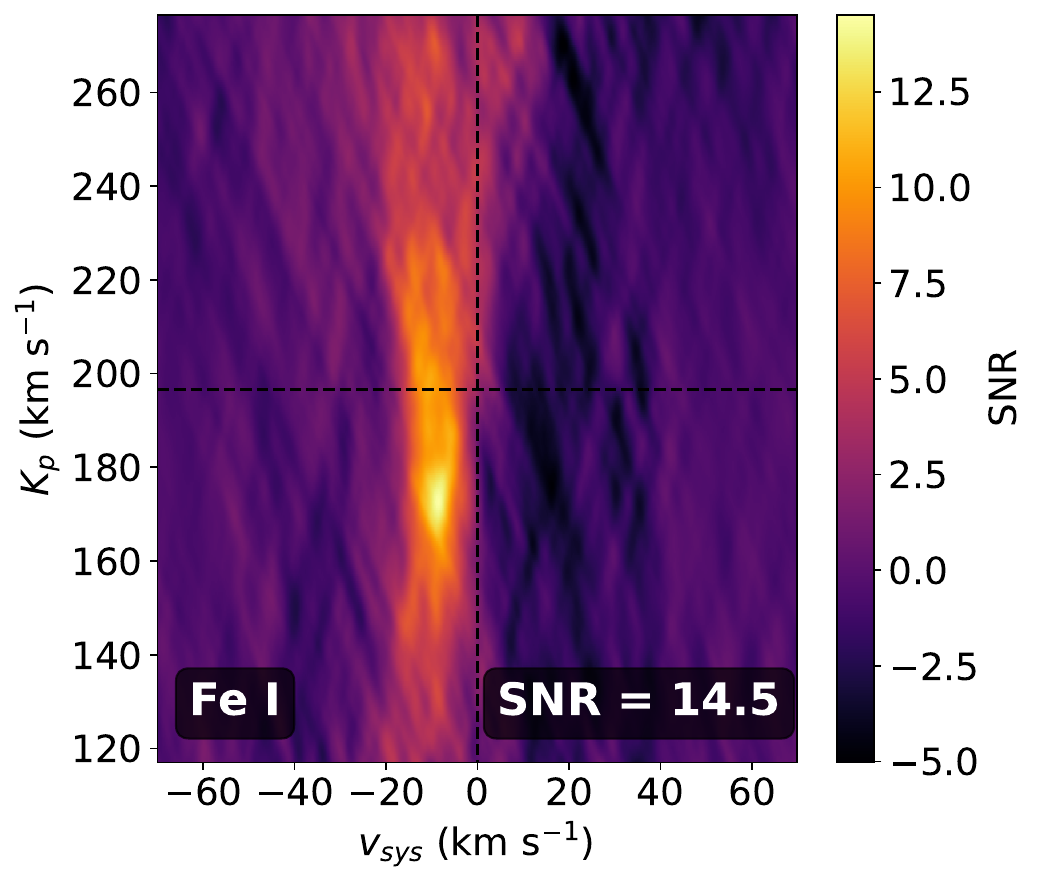}
    \includegraphics[width=0.32\linewidth]{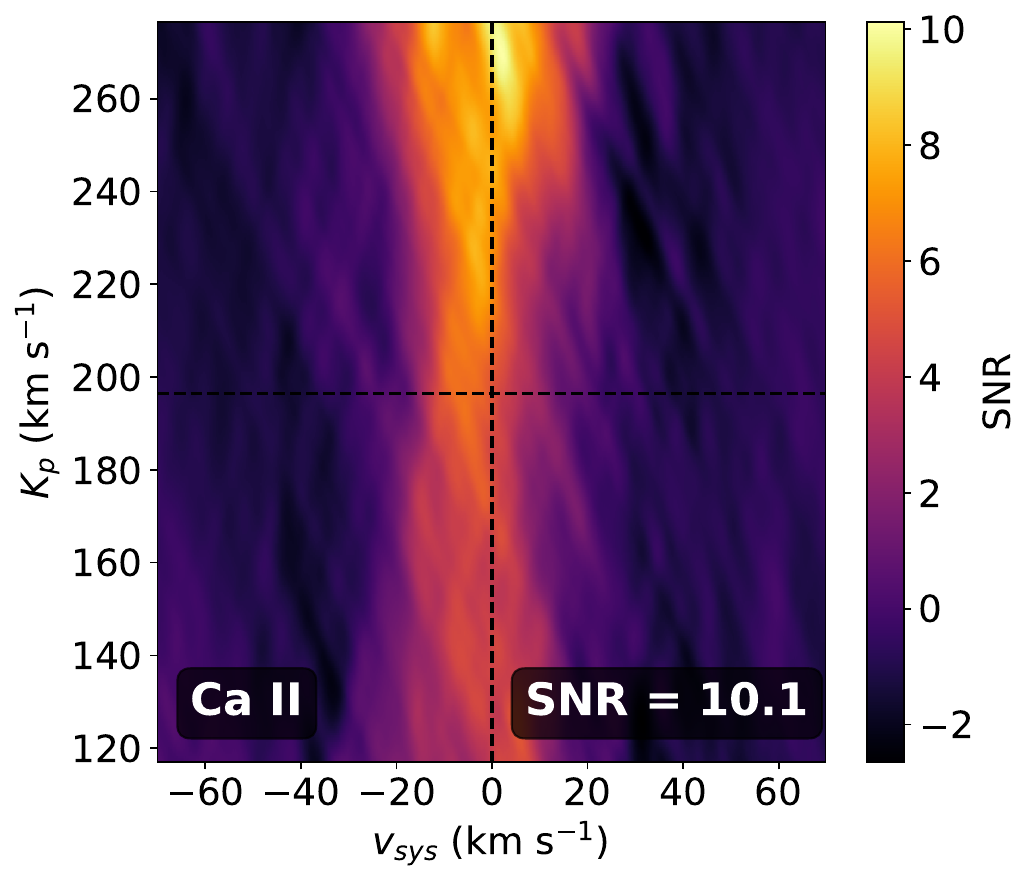}
    \includegraphics[width=0.32\linewidth]{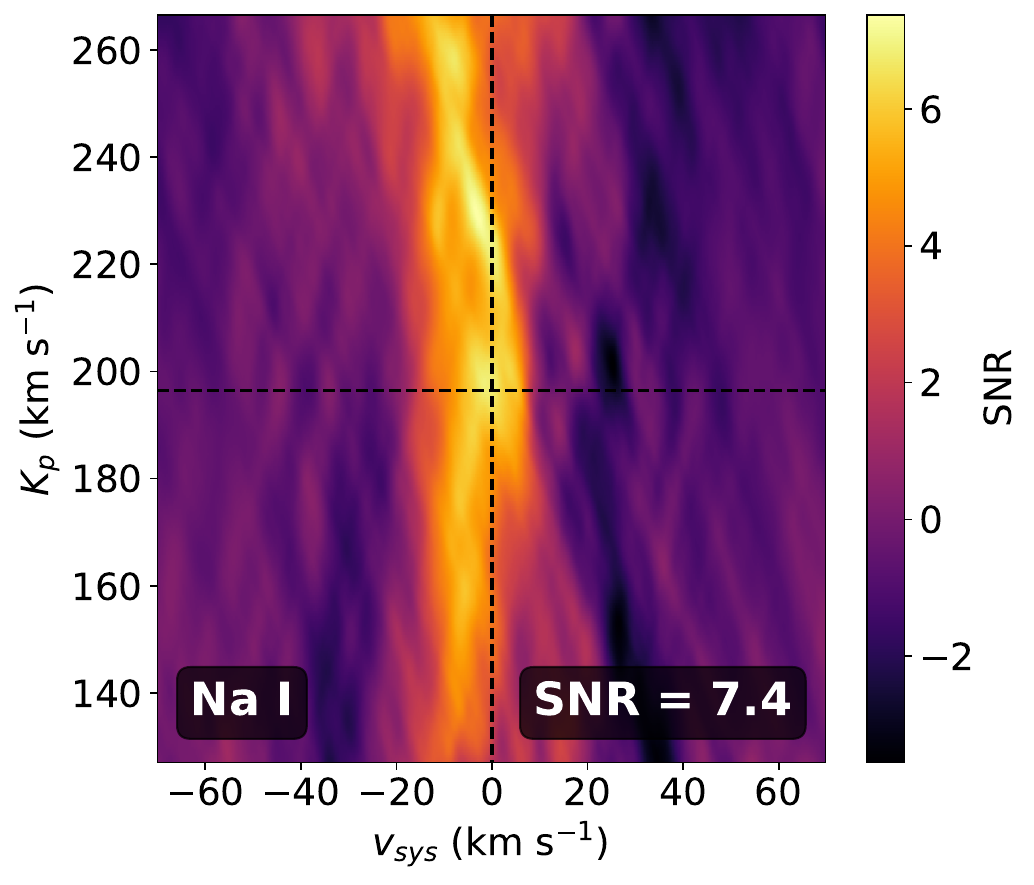}
    \includegraphics[width=0.32\linewidth]{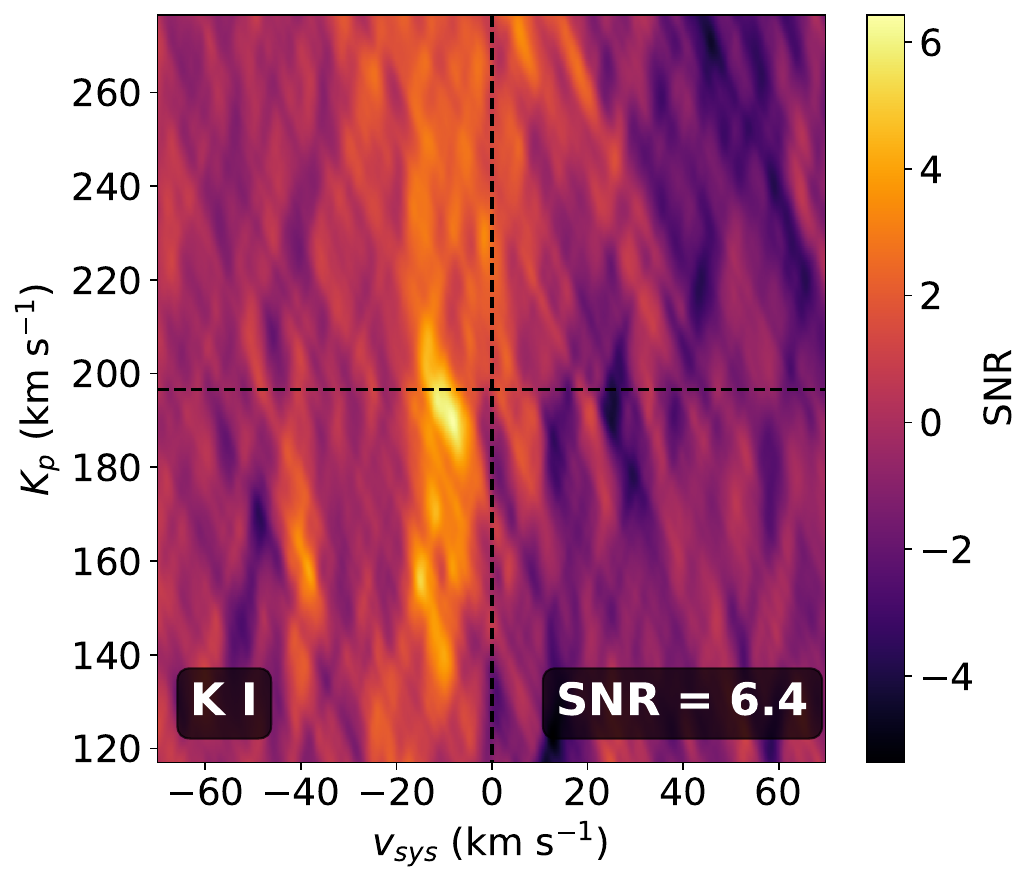}
    \includegraphics[width=0.32\linewidth]{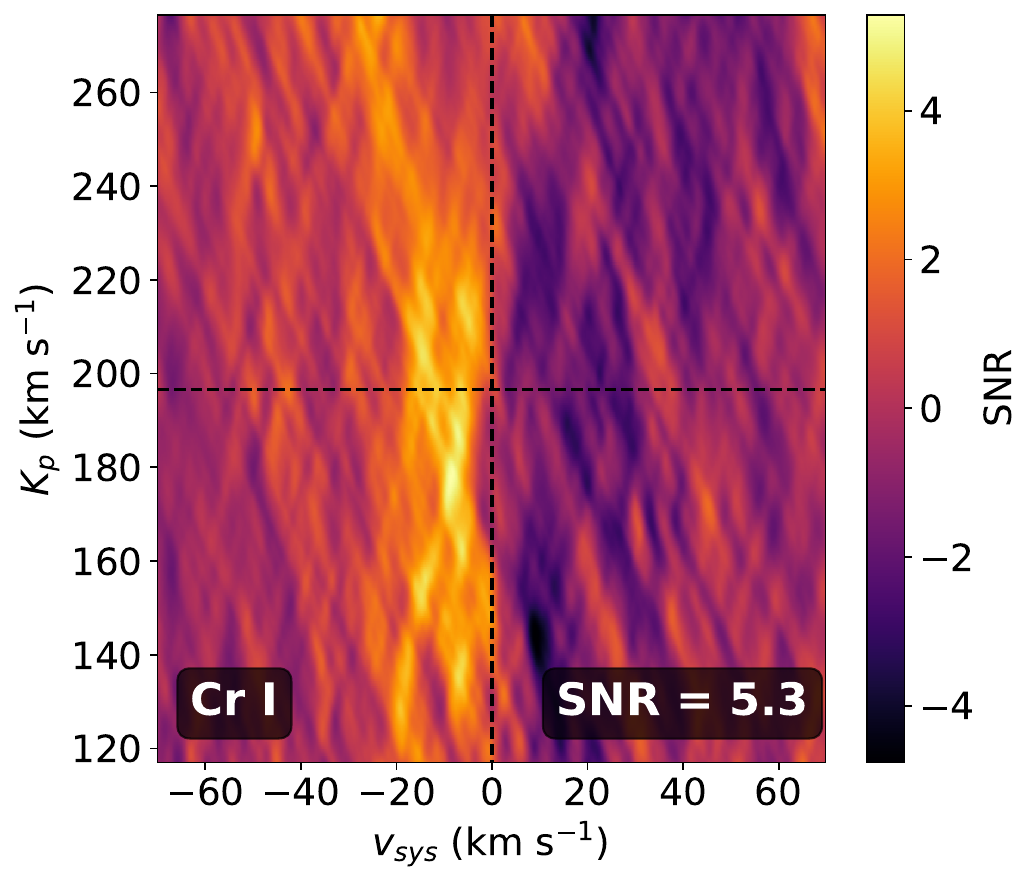}
    \includegraphics[width=0.32\linewidth]{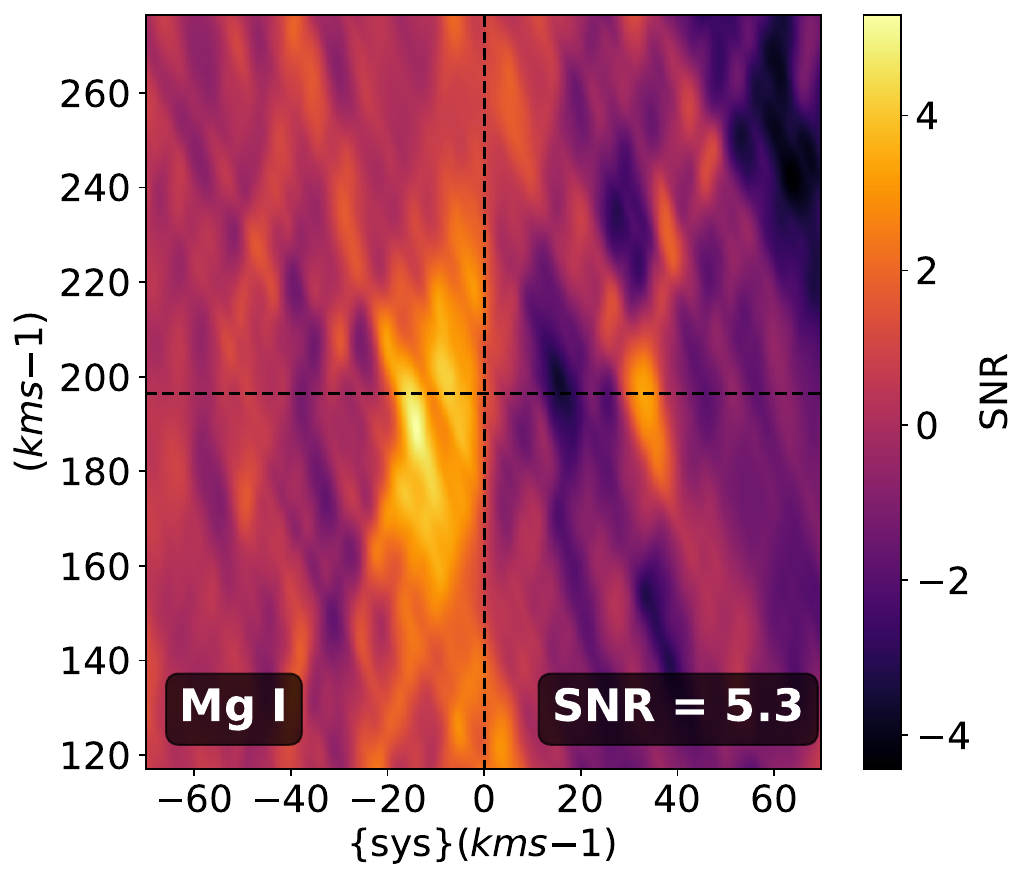}
    \includegraphics[width=0.32\linewidth]{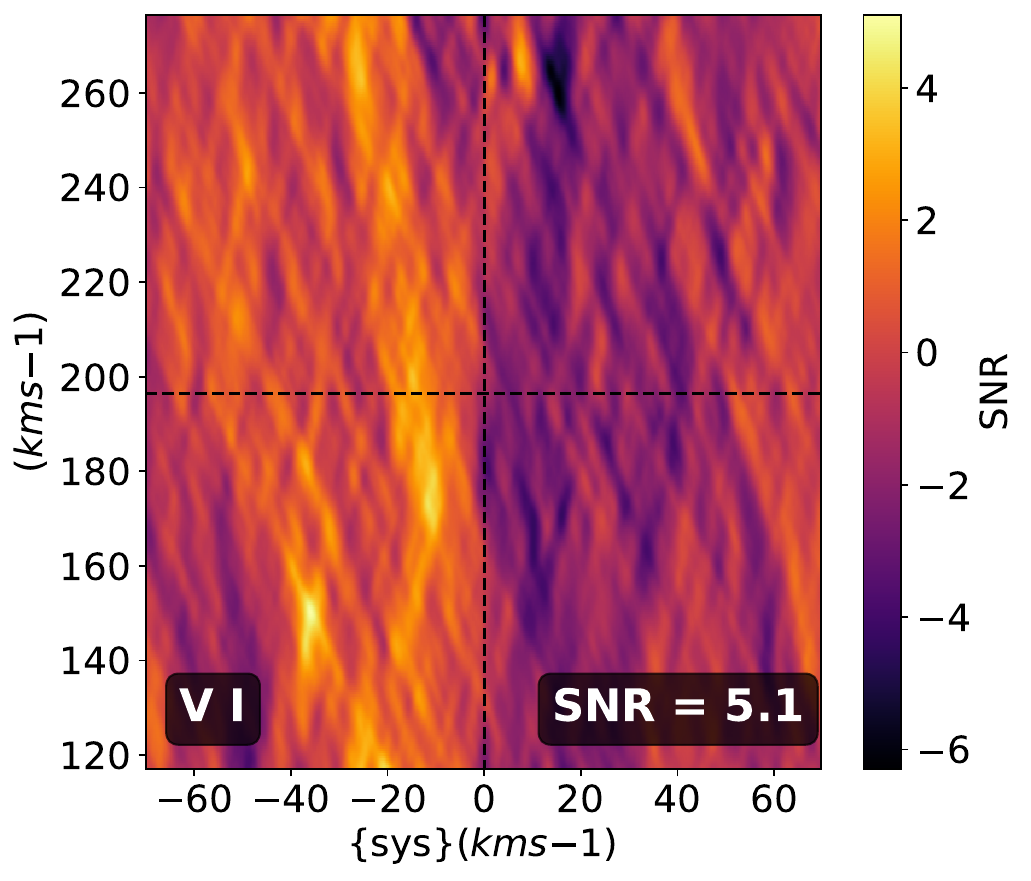}
    \includegraphics[width=0.32\linewidth]{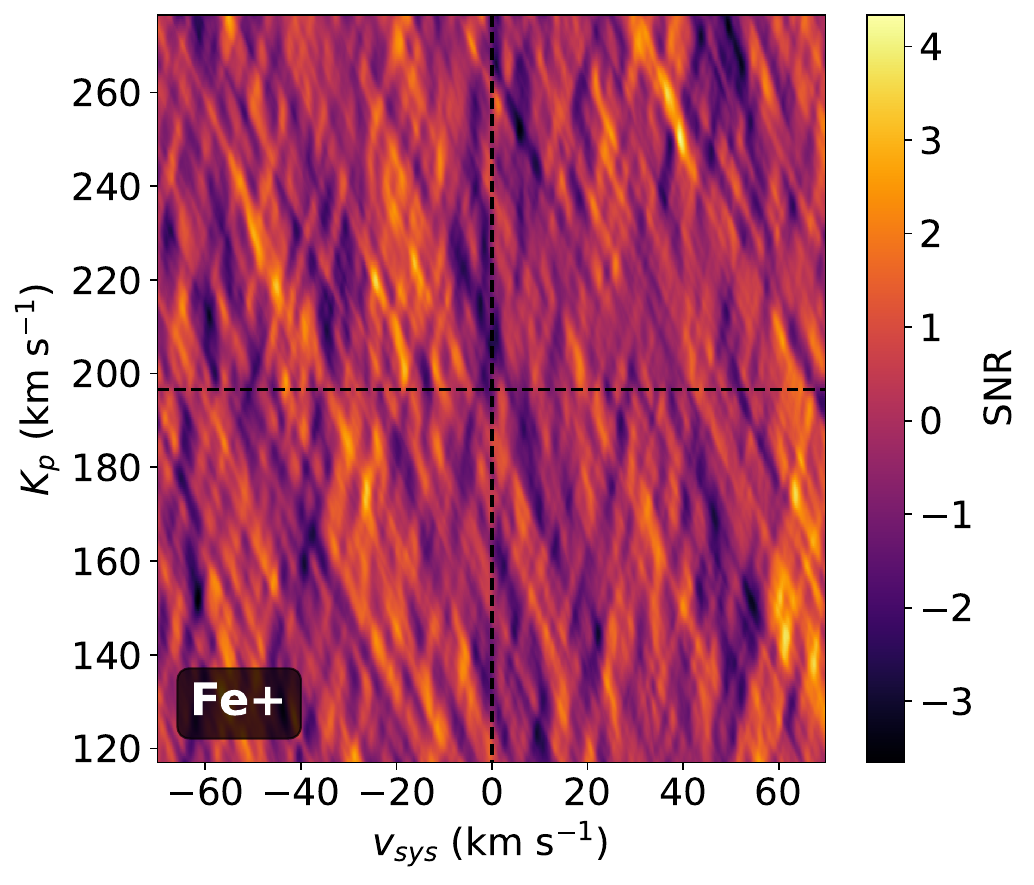}
    \includegraphics[width=0.32\linewidth]{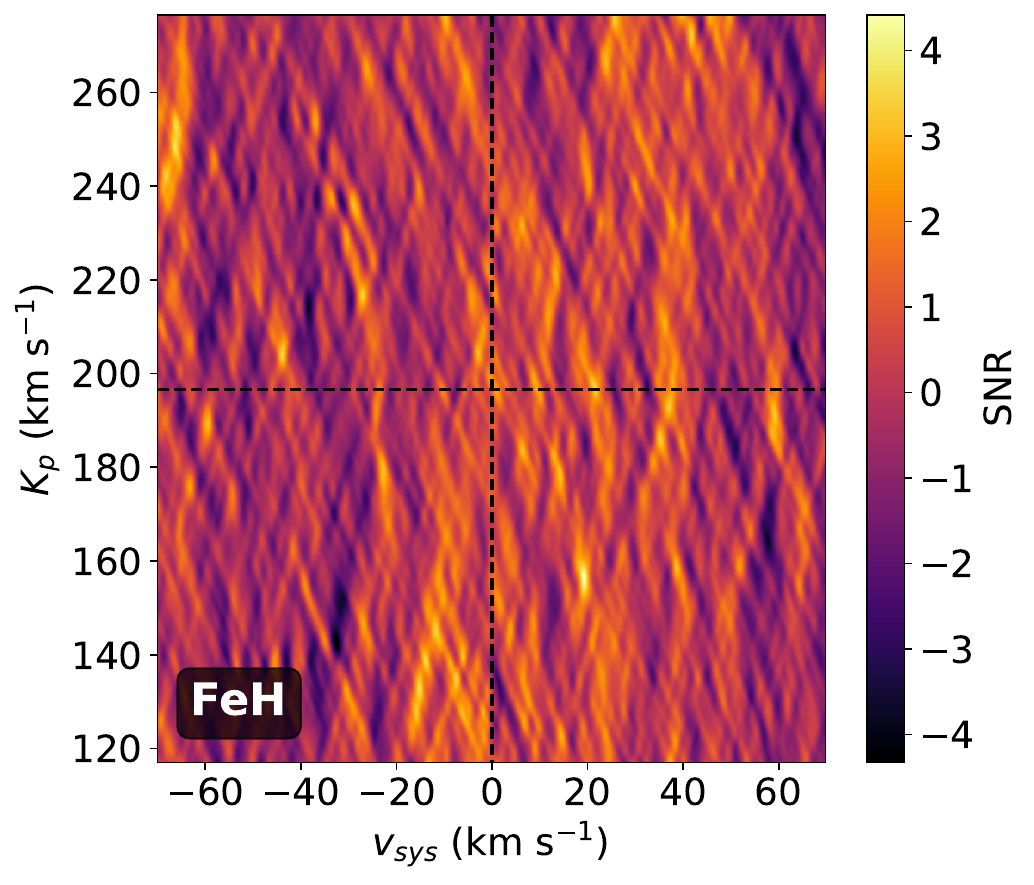}
    \includegraphics[width=0.32\linewidth]{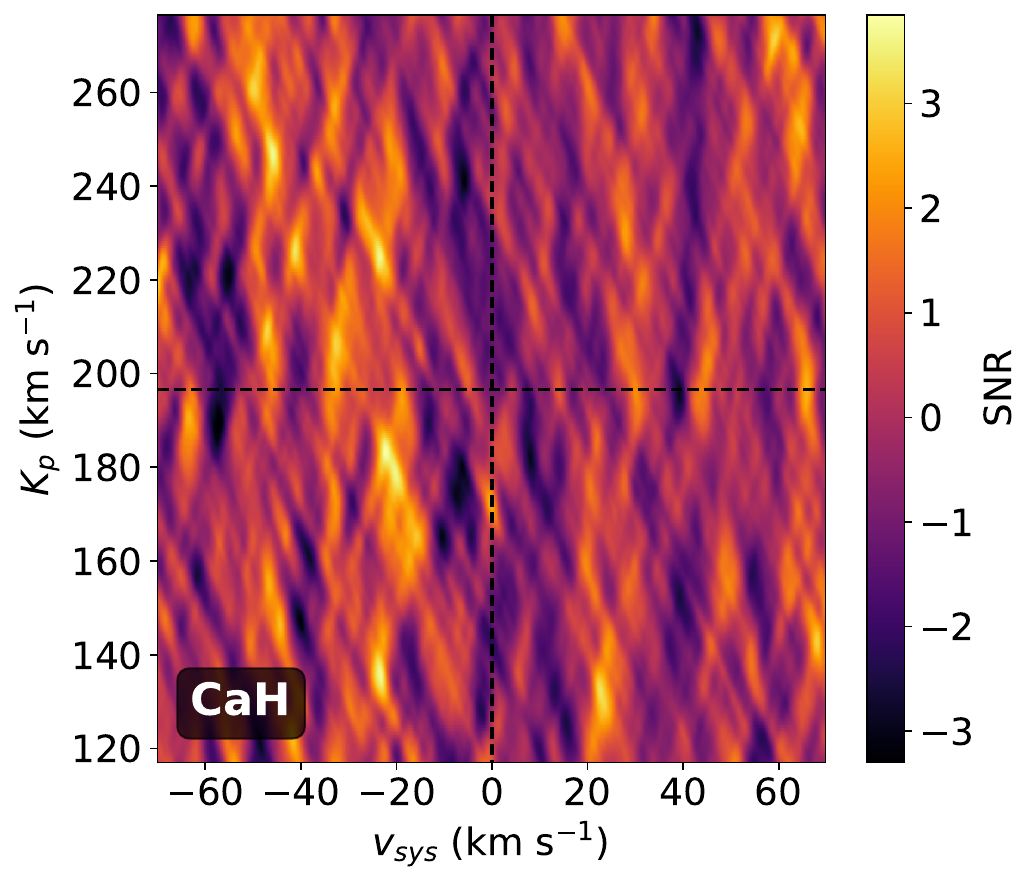}
    \includegraphics[width=0.32\linewidth]{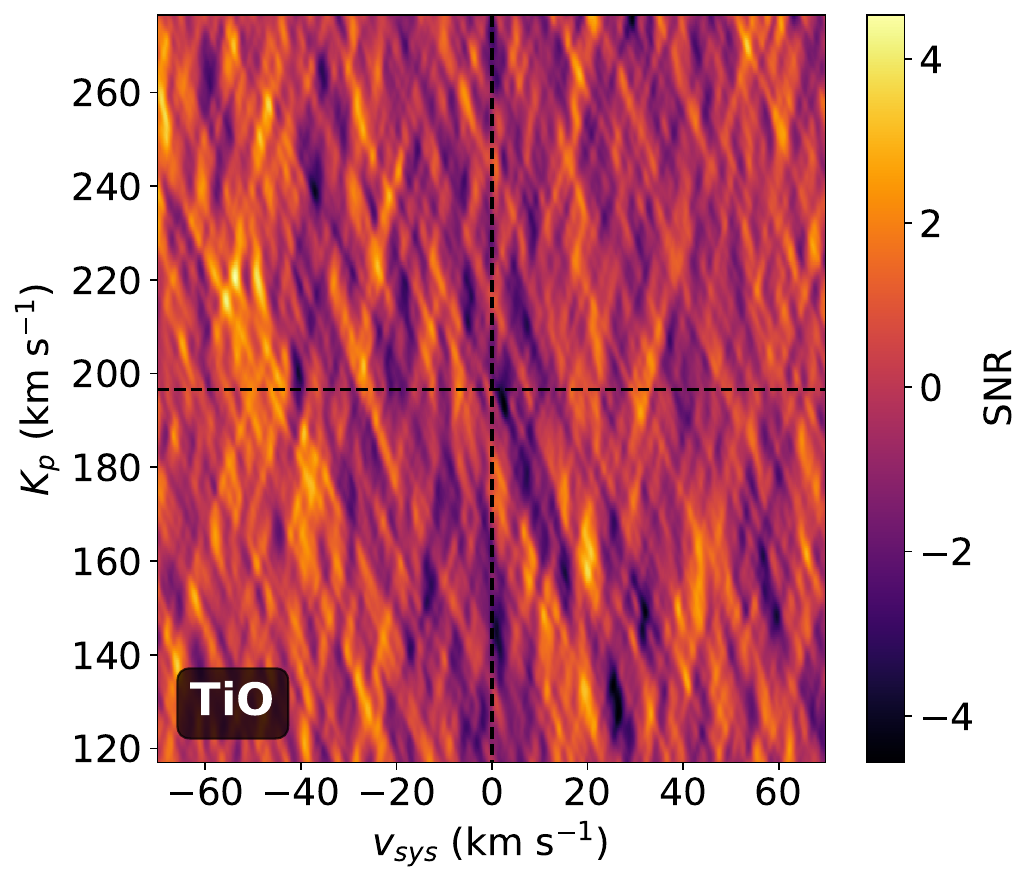}
    \includegraphics[width=0.32\linewidth]{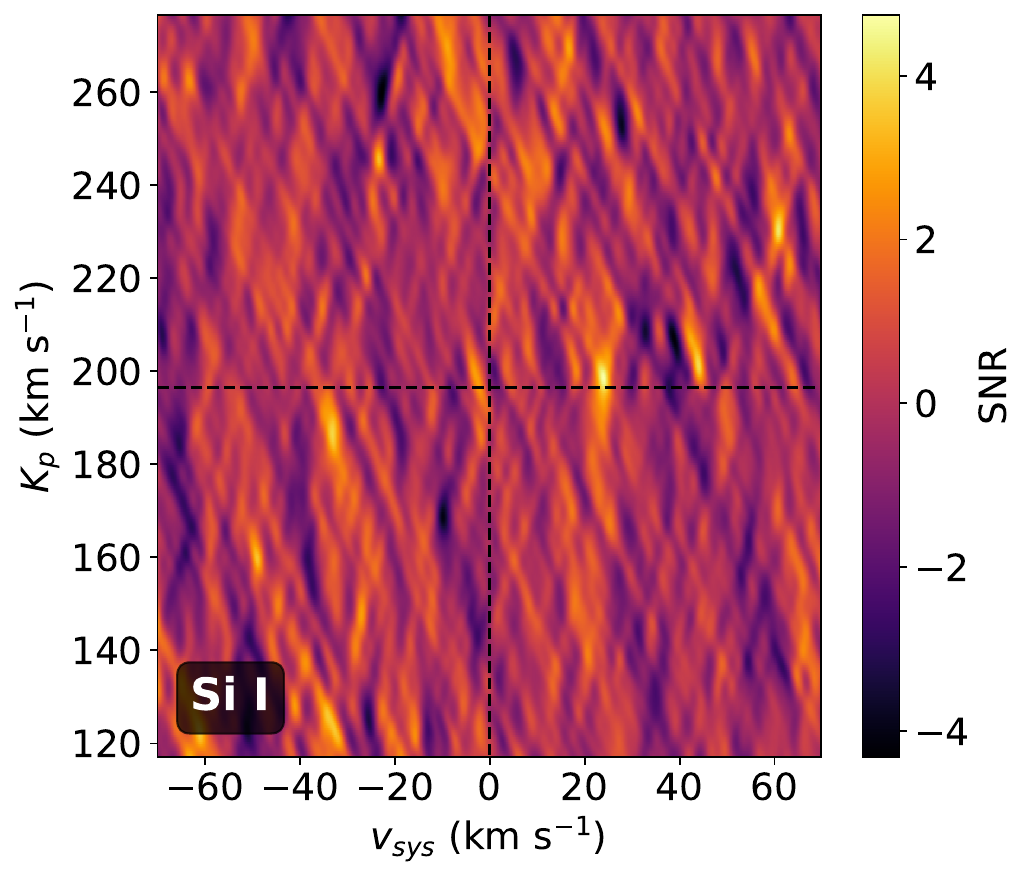}
    \caption{Two–dimensional $K_{p}$–$v_{\rm sys}$ map showing the detection significance of various species. \red{For Fe I, Ca II, and Na I, the large yellow residuals near the expected planetary velocity (black dashed lines) indicate strong atmospheric absorption signals. K I, Cr I, V I, and Mg I show weaker signals, while Si I, Fe II, FeH, CaH, and TiO are not detected.} We caution that these SNR values should not be interpreted as true $\sigma$-significances, as doing so may overestimate the confidence of the detections \citep{Cabot2019,Kipping2025}.}
    \label{fig6}  
\end{figure*}
After generating the two-dimensional cross-correlation functions for multiple species, we analyzed the in-transit residuals to isolate the planetary absorption signal. Notably, the absorption features in Figure \ref{fig:ccmap} are not centered at 0 km s$^{-1}$ as expected in the absence of atmospheric dynamics but are blue-shifted by several km s$^{-1}$. Such a blue-shift is a well-established signature of day-to-night winds in hot Jupiter atmospheres \citep{Showman2002,Snellen2010,Wyttenbach2015}. Specifically, because WASP-76 b is likely tidally locked, its permanent day-side receives intense stellar irradiation, driving strong winds from the hotter day-side to the cooler night-side. On the leading and/or trailing limbs, these winds can have a line-of-sight velocity toward the observer during transit, resulting in a blue-shift of the planetary absorption signal in the planet rest frame.

A more striking feature of the cross-correlation plots in Figure \ref{fig:ccmap} is that there is a strong ingress–egress asymmetry for some species (e.g., Fe I) but not others (e.g., Ca II). To quantify this behavior, we co-added the in-transit CCFs in two phase windows, $\phi\in[-0.038,-0.020]$ (near ingress) and $\phi\in[+0.020,+0.038]$ (near egress), and fit a single Gaussian to each co-added CCF. This phase-binning helped to stabilize the Gaussian fits that were otherwise unreliable on individual phases at lower SNR. Similar strategies have been demonstrated to be effective for resolving atmospheric asymmetries in high-resolution spectroscopy \citep[e.g.,][]{Savel2023}; however, we caution that the stacked CCF profiles appear to be highly non-Gaussian (particularly for Na I and Ca II), which made it difficult to estimate line centers and uncertainties with high fidelity. In practice, we found that reasonable changes to the fitting method could shift the recovered line centers of Na I and Ca II by up to a few km s$^{-1}$. Nonetheless, the qualitative behavior we observed was highly consistent across all different fitting techniques: Fe I shows clear phase-dependent asymmetries consistent with day-to-night winds (ingress $=$ -4.3 km s$^{-1}$ and egress $=$ -11.3 km s$^{-1}$), while Na I and Ca II behave differently with no measurable ingress–egress asymmetry (Na I ingress $=$ -4.9 km s$^{-1}$ and egress $=$ -3.0 km s$^{-1}$; Ca II ingress $=$ -2.1 km s$^{-1}$ and egress $=$ -2.6 km s$^{-1}$). We also find signs of an asymmetry in Cr I (near ingress $=$ -4.2 km s$^{-1}$ and egress $=$ -10.9 km s$^{-1}$), though at much lower significances, given the weaker signal near ingress (Figure \ref{fig:ccmap}). 

In addition to the cross-correlation residual plots, we also chose to quantify detection significances using the standard two-dimensional $K_p$–$v_{\rm sys}$ plots \citep{Brogi2012}.  Assuming a circular orbit, we calculated the instantaneous planetary radial velocity using $v_p(t) = K_p \sin(2\pi \phi(t)) + v_\mathrm{sys}$, where $\phi(t)$ is the orbital phase. For each trial pair of \textit{$K_p$} and \textit{$v_\mathrm{sys}$}, we shifted the in-transit CCFs by $-v_p(t)$ to align the expected planetary signal at 0 km s$^{-1}$. To evaluate the significance of the detection at each point, we computed the SNR by dividing the combined CCF value at zero velocity by the standard deviation of the CCF values more than 50 km s$^{-1}$ from the expected signal location. In this framework, strong detections appear as bright localized peaks centered close to the known $K_p$ and $v_{\rm sys}$ values (Figure \ref{fig6}). \red{We strongly detect Fe I, Ca II, and Na I, with phase-resolved cross-correlation features visible across both ingress and egress (Figure~\ref{fig:ccmap}). At lower significance, we also find signals from K I, Cr I, V I, and Mg I (Figure~\ref{fig6}), although these species are not as strongly phase resolved. An example of the weaker phase-resolved structure is shown for Cr I in Figure~\ref{fig:ccmap}. The remaining species tested (Si I, Fe II, FeH, CaH, TiO) show no significant signal near the expected \textit{$K_p$} and \textit{$v_\mathrm{sys}$}} (Figure \ref{fig6}).

\section{Discussion}
\subsection{Comparison to previous studies on WASP-76 b}
\label{discussion}
The ingress–egress asymmetry in Fe I for WASP-76 b has been extensively studied across multiple papers and instruments \citep{Ehrenreich2020, Kesseli2021, Kesseli2022, pelletier2023}. \citet{Ehrenreich2020} first interpreted this phase-dependent blue-shift as possible evidence for condensation of Fe I on the night-side of WASP-76 b. However, \citet{Kesseli2022} and \cite{pelletier2023} found the same effect in multiple species, suggesting a bulk atmospheric process rather than condensation. Condensation would most likely deplete a single species rather than produce coherent velocity shifts across several independent absorbers, whereas bulk atmospheric processes could affect many species simultaneously. 

Many papers have modeled the time-varying Fe I signal on WASP-76 b with GCMs in order to explain this blue-shift evolution \citep{Wardenier2021, Savel2022}. The emerging picture from these models is that a day-to-night wind combined with a cooler western (morning) limb can qualitatively explain the increasing blue-shift from ingress to egress. However, the GCMs struggle to match both the magnitude and the phase evolution of the signal without additional physics. In particular, forward models show that reproducing large ($\gtrsim$ 5 km s$^{-1}$) blue-shifts across transit requires weak drag and a deep radiative–convective boundary \citep{Wardenier2021, Savel2022}. However, even these GCMs only reproduce the observed blue-shift evolution when optically thick clouds or slight orbital eccentricity are included, which suggests that there may be missing physics in the GCM models. Interestingly, despite these modeling challenges, the measured Fe I blue-shift itself remains relatively stable from the first HARPS-North transit (taken in 2012) to our KPF transit (taken in 2023) at $\sim$6$-$9 km s$^{-1}$ blueshift evolution from ingress to egress \citep{Kesseli2021, Kesseli2022}. The consistency across instruments and epochs suggests that the underlying mechanisms responsible for the asymmetric absorption do not lead to strong variability on decade-long timescales. Nonetheless, continued monitoring with KPF and other ultra-stable spectrographs will be essential for determining whether subtle epoch-to-epoch variability exists and for isolating the physical mechanisms that lead to the strong asymmetric absorption in Fe I.

\red{In our single KPF transit, Fe I is the only species for which we confidently measure an ingress-to-egress velocity asymmetry. Cr I shows a hint of similar behavior, and this interpretation is supported by the Cr I asymmetry reported by \citet{Kesseli2022}, but the ingress signal in our data is too weak to claim a phase-resolved asymmetry with confidence. By contrast, Na I and Ca II are both detected strongly enough to measure phase-resolved velocities, but neither species shows a measurable ingress-to-egress asymmetry in our data. For Na I, this is consistent with \citet{Kesseli2022}, who also found very little radial velocity difference between ingress and egress ($-3.8 \pm 0.8$ to $-6.1 \pm 1.5$ km~s$^{-1}$). Our KPF transit gives similarly symmetric Na I velocities (ingress $= -4.9$ km~s$^{-1}$, egress $= -3.0$ km~s$^{-1}$). For Ca II, \citet{Kesseli2022} reported ingress-to-egress RV values of $+8.1 \pm 2.3$ to $-2.5 \pm 1.5$ km~s$^{-1}$, but noted that the Ca II CCF was double-peaked and difficult to characterize. Our phase-resolved Ca II CCF is also broad, but remains nearly vertical in velocity space, with no measurable ingress-to-egress difference (ingress $= -2.1$ km~s$^{-1}$, egress $= -2.6$ km~s$^{-1}$).}

\begin{figure}
    \centering
    \includegraphics[width=\linewidth]{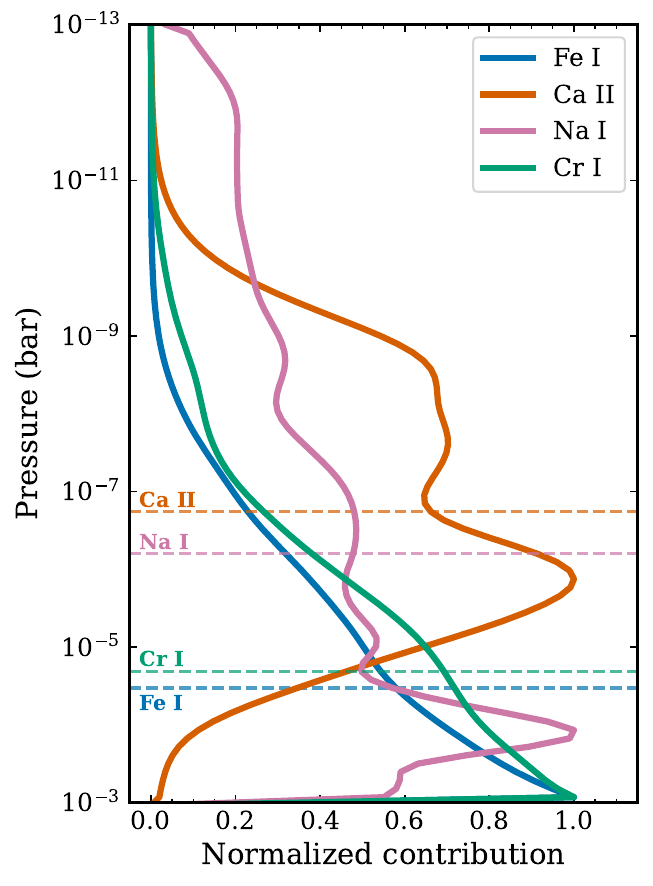}
    \caption{Velocity-weighted contribution functions for Fe I (blue), Ca II (orange), Na I (pink), and Cr I (teal), computed using \texttt{petitRADTRANS} with an isothermal P-T profile at 3000 K. For each species, the two-dimensional contribution function $C(P, \lambda)$ was weighted by the \citet{Bouchy2001} velocity information weight $W(\lambda)$ (Equation~\ref{eq:bouchy}) and summed over wavelength, yielding the pressure distribution that drives the velocity signal recovered by the cross-correlation. Contributions from pressures greater than $10^{-3}$ bar are excluded, as these deep layers are optically thick and do not meaningfully contribute to the transit signal. Dashed horizontal lines mark the median effective pressure for each species: $1.8\times10^{-7}$ bar for Ca II, $6.3\times10^{-7}$ bar for Na I, $2.0\times10^{-5}$ bar for Cr I, and $3.4\times10^{-5}$ bar for Fe I. Because lower pressures correspond to higher altitudes, Ca II and Na I probe significantly higher in the atmosphere than Fe I and Cr I, which trace deeper layers where atmospheric dynamics differ.}
    \label{fig:contribution}
\end{figure}

\red{The picture that emerges from our KPF data is therefore that the velocity behavior differs from species to species. This species-dependent behavior is consistent with the idea that the measured velocities trace different atmospheric layers, since previous modeling and observations of WASP-76 b show that dynamical signatures can vary with pressure \citep{Wardenier2021,Savel2022,Kesseli2024}. We therefore compute velocity-weighted contribution functions to identify which pressures dominate the cross-correlation signal for each species.} 
\red{Because the cross-correlation measures a Doppler shift, the relevant quantity is not which pressures contribute to the transit depth at each wavelength, but which pressures contribute to the wavelengths that carry the most velocity information. Using the \texttt{petitRADTRANS} templates described in Section~\ref{sec4.1}, we first obtained the two-dimensional contribution function $C(P, \lambda)$, which measures how strongly each pressure layer contributes to the effective transit radius at each wavelength. We then weighted $C(P, \lambda)$ by the velocity information content of each wavelength pixel, adopting the weighting scheme of \citet{Bouchy2001},}
\begin{equation}
    W(\lambda) \propto \lambda^2
    \left(\frac{\partial A_0}{\partial \lambda}\right)^2
    \frac{1}{A_0(\lambda)},
    \label{eq:bouchy}
\end{equation}
\red{where $A_0(\lambda)$ is the model transmission spectrum convolved to the KPF resolution of $R = 97{,}000$. Summing $W(\lambda)\,C(P,\lambda)$ over wavelength yields the pressure distribution that drives the velocity signal recovered by the cross-correlation for each species (Figure~\ref{fig:contribution}). We exclude contributions from pressures greater than $10^{-3}$ bar, as \citet{Kesseli2024} found that Fe lines below this pressure level do not meaningfully contribute to the cross-correlation signal, consistent with the cloud deck location retrieved by \citet{Gandhi2022}.}

\red{As shown in Figure~\ref{fig:contribution}, the median effective pressures for Ca II and Na I are $1.8\times10^{-7}$ and $6.3\times10^{-7}$ bar, respectively, while Fe I and Cr I probe significantly deeper layers with medians of $3.4\times10^{-5}$ and $2.0\times10^{-5}$ bar. In the higher-altitude (lower-pressure) layers probed by Ca II and Na I, both the leading and trailing limbs may remain optically thick over a larger fraction of the transit, so their respective red-shifted and blue-shifted velocity components both contribute to the observed signal at most transit phases. In contrast, Fe I probes deeper layers where the trailing limb may contribute more strongly at later phases of transit, producing the larger blue-shift and asymmetric absorption observed for Fe I. This interpretation is qualitatively consistent with previous GCM modeling of WASP-76 b that found that when the Fe I signal is decomposed by altitude, the magnitude of the blue-shift decreased systematically at higher atmospheric layers \citep{Kesseli2024}. The same species-dependent velocity behavior is also visible in the $K_p$--$v_{\rm sys}$ maps, where Fe I and Cr I show peak velocities lower than the expected $K_p$, while Ca II and Na I show peak signals above the expected $K_p$.}

\red{
\subsection{Comparison to WASP-121 b}
Interestingly, our results for WASP-76 b significantly differ from those recently reported for WASP-121 b \citep{Seidel2025}. While Fe I shows a blue-shift in WASP-121 b comparable to our observations in WASP-76 b, the higher-altitude tracers in WASP-121 b (e.g., Na I and H$\alpha$) show dramatically larger blue-shift evolution, with the Na I blue-shift increasing from ingress to egress by nearly 40 km s$^{-1}$. This behavior was interpreted as evidence for a high-altitude super-rotating jet in the atmosphere of WASP-121 b. However, such jets are not expected to dominate at the high altitudes probed by H$\alpha$ or Na I according to most general circulation models (e.g., \citealt{Showman2011}; see also Figure 8 of \citealt{Snellen2025}). Indeed, \citet{Seidel2025} note that this behavior is difficult to reconcile with current GCM predictions.
Therefore, unlike WASP-121 b, our observations of WASP-76 b do not require invoking an additional high-altitude circulation component beyond those predicted by current GCMs. These differences between WASP-76 b and WASP-121 b suggest that atmospheric circulation patterns in ultra-hot Jupiters may vary significantly between planets in ways not currently predicted by existing GCMs. Fully understanding the origin of these differences will likely require both improved GCMs and additional high-resolution observations across a broader range of planetary and stellar properties.}

\section{Conclusion}
\label{conclusion}
In this paper, we have introduced the KPF SURFS-UP Survey, a new high-resolution effort to study the atmospheres of UHJs with the recently commissioned KPF spectrograph. We developed a comprehensive pipeline tailored to reducing KPF data, which includes blaze removal, continuum normalization, science spectra combination, telluric correction, and atmospheric detection. Though some user input is required (e.g., the inclusion and exclusion regions for \texttt{Molecfit}), this pipeline is otherwise fully automated, so it can be quickly applied to many datasets across a wide variety of science cases. For instance, our pipeline will also be useful for measuring stellar abundances with KPF. In the future, we hope to fully automate all steps in the pipeline so that it can simply run entirely by changing the input files as well as the stellar and planetary parameters.

As a first demonstration, we applied this pipeline to a transit observation of WASP-76 b, performing transmission spectroscopy via cross-correlation. The main findings on WASP-76 b are as follows:
\begin{itemize}
\item \red{We strongly detect Fe I, Ca II, and Na I from a single KPF transit of WASP-76 b, with phase-resolved cross-correlation features visible across both ingress and egress. At lower significance, we find signals from K I, Cr I, V I, and Mg I, though without the strong phase-resolved signatures seen in Fe I, Ca II, and Na I. The remaining species tested (Si I, Fe II, FeH, CaH, TiO) are not detected.} Fe I was detected at an SNR $\sim$14.5, \red{comparable to the Fe I detection significance reported by \citet{Kesseli2022} using two transits with ESPRESSO,} establishing KPF as an optical workhorse for exoplanet atmospheres.
\item We measured strong phase-dependent blue-shifts from ingress to egress for Fe I, as well as signs of an asymmetry for Cr I. The magnitude of the blue-shift we measured for Fe I is highly consistent with past ESPRESSO and HARPS results, indicating that the asymmetry is relatively stable over several-year timescales.

\item In contrast to Fe I, Ca II and Na I show no measurable ingress–egress asymmetry and remain nearly vertical in velocity space, consistent with absorption at higher altitudes where the asymmetry between ingress and egress is weaker than in the deeper layers traced by Fe I.

\end{itemize}

The success of these initial observations highlights the potential of using KPF for studying UHJ atmospheres as well as the atmospheres of smaller planets. Looking ahead, future papers in this survey will feature observations of over a dozen UHJs using transmission, emission, and phase-resolved spectroscopy. With this large sample of planets, we will be able to constrain the refractory abundances of UHJs on a population level, which may provide important insights into hot Jupiter formation and evolution. Furthermore, these observations will also allow us to test predictions from general circulation models, which will improve our understanding of the complex three-dimensional nature of exoplanet atmospheres. 

\section*{Acknowledgments}
This material is based upon work supported by the National Science Foundation Graduate Research Fellowship under Grant No. 2141064 and the MIT Dean of Science Fellowship. We also acknowledge the use of computational resources provided by the MIT Engaging cluster at the Massachusetts Green High Performance Computing Center (MGHPCC). L.M.W. acknowledges support from the NASA Exoplanet Research Program (grant no. 80NSSC23K0269). A.B.-A.'s contribution to this research was carried out at the Jet Propulsion Laboratory, California Institute of Technology, under a contract with the National Aeronautics and Space Administration (80NM0018D0004).

This research has made use of the NASA Exoplanet Archive, which is operated by the California Institute of Technology, under contract with the National Aeronautics and Space Administration under the Exoplanet Exploration Program \citep{Akeson2013,Christiansen2025}. Some of the data presented herein were obtained at Keck Observatory, which is a private 501(c)3 non-profit organization operated as a scientific partnership among the California Institute of Technology, the University of California, and the National Aeronautics and Space Administration. The Observatory was made possible by the generous financial support of the W. M. Keck Foundation. 

The authors wish to recognize and acknowledge the very significant cultural role and reverence that the summit of Maunakea has always had within the Native Hawaiian community. We are most fortunate to have the opportunity to conduct observations from this mountain.

\bibliography{sample7}{}
\bibliographystyle{aasjournalv7}

\end{document}